\documentclass[twocolumn]{aastex631}

\usepackage{xcolor}
\usepackage{etoolbox}

\usepackage{graphicx}
\usepackage{amsmath}
\usepackage{amsfonts}

\begin{document}

\title{Fast Magnetoacoustic Wave Behavior within Gravitationally Stratified, Magnetically Inhomogeneous Media}

\author[0009-0003-4401-1027]{Ryan T. Smith}
\affiliation{Northumbria University, Newcastle upon Tyne, NE1 8ST, UK}

\author[0000-0002-7863-624X]{James A. McLaughlin}
\affiliation{Northumbria University, Newcastle upon Tyne, NE1 8ST, UK}

\author[0000-0002-5915-697X]{Gert J. J. Botha}
\affiliation{Northumbria University, Newcastle upon Tyne, NE1 8ST, UK}



\begin{abstract}
\noindent
The nature of MHD waves within inhomogeneous media is fundamental to understanding and interpreting wave behavior in the solar atmosphere. We investigate fast magnetoacoustic wave behavior within gravitationally stratified, magnetically inhomogeneous media, by studying a magnetic environment containing a simple 2D X-type magnetic null point. The addition of gravitational stratification fundamentally changes the nature of the system, including breaking the symmetry. There are two main governing effects: the stratified density profile acts in combination with the inhomogeneous magnetic field, creating a large gradient in the Alfv\'en speed and hence a system replete with refraction. The system is investigated using both numerical simulations and a semianalytical WKB solution (via Charpit’s method and a fourth-order Runge-Kutta solver) and we find strong agreement between both. The results show a fundamental difference between the stratification-free and stratified cases, including the formation of caustic surfaces and cusps, and we contextualize these results in the theoretical understanding of fast magnetoacoustic waves. 
\end{abstract}




\section{Introduction} \label{sec:intro}

The solar corona is a complex system which is characterized by highly magnetized, low-density plasma, playing host to intricate magnetic field structures. Despite its high temperatures, the corona is modeled as a low-$\beta$ plasma, as it is magnetically dominated. Observations of wave motions in the solar corona have been reported in a number of works, such as \cite{aschwanden1999coronal} and \cite{nakariakov1999trace}, and reviewed more recently by \cite{nakariakov2020magnetohydrodynamic}, \cite{banerjee2021magnetohydrodynamic}, and \cite{morton2023alfvenic}, with these review papers considering a variety of MHD wave modes. 

The introduction of gravity to an MHD system (such as when modeling the solar corona) has several important consequences \citep{roberts2019mhd}. Any MHD equilibrium must remain in force balance, and combining this property with the presence of gravity leads to gravitational stratification. This stratification introduces a directional dependence to the medium (in the same way that the introduction of a magnetic field leads to structuring). The presence of gravity also introduces a length scale into the medium, namely the scale height. The inclusion of gravity gives rise to a variety of different oscillations, such as internal gravity waves (IGWs) \citep{hague2016magnetohydrodynamic,roberts2019mhd}.

The propagation of MHD waves in a stratified atmosphere has been investigated in a variety of studies, with one example being \cite{de2004damping}, who considered the propagation of the slow magnetoacoustic wave along coronal loops and the effect gravitational stratification has on the propagation. It was found that gravitational stratification causes the amplitude of the wave to increase as it propagates upward away from the photosphere, which is a consequence of the decreasing background density. Slow magnetoacoustic waves also experience a cutoff frequency (see the review by \cite{roberts2019mhd}). \cite{ferraro1958hydromagnetic} derived the equations governing magnetoacoustic wave behaviour in a purely uniform, vertical magnetic field, in a system with an invariant direction. They found that the Alfv\'en waves are decoupled from the slow and fast magnetoacoustic waves. \cite{bogdan2003waves} investigated magnetoacoustic wave behavior in a nonuniform, 2D stratified magnetoatmosphere, reporting upon the evolution of both slow and fast magnetoacoustic waves, as well as their coupling where the Alfv\'en and sound speeds are equal. \cite{gao2024propagating} investigated the propagation of the kink wave in a coronal magnetic flux tube and considered the effects of resonant absorption and gravitational stratification on the altitude variation of the wave amplitude. \cite{article} investigated an equivalent system to that of \cite{bogdan2003waves} and deployed a novel wave mode identification technique to identify the slow and fast modes in simulation.

Potential field extrapolations of photospheric magnetograms have shown that magnetic null points are ubiquitous within the solar corona, as seen in \cite{brown2001topological}, \cite{longcope2009number}, and \cite{regnier2013magnetic}. These \lq{weak}\rq{} points in the coronal magnetic field are locations at which the magnetic field strength, and consequently the Alfv\'en speed, are zero and as a result of this magnetic null points become an important feature when considering the propagation of MHD waves, in particular the fast magnetoacoustic wave which is investigated in this paper.

The propagation of MHD waves around magnetic null points has been the subject of a number of studies. For example, \cite{mclaughlin2004mhd} investigated the propagation of the linear fast magnetoacoustic and Alfv\'en waves in the neighborhood of a 2D X-type magnetic null point under the low-$\beta$ approximation finding that the linear fast magnetoacoustic wave wraps around the null point, with the wave becoming trapped close to, but never reaching, the null point. \cite{mclaughlin2005mhd} extended that investigation to a magnetic topology containing two null points, removing the simple radial symmetry from \cite{mclaughlin2004mhd}. Similar behavior was observed, though the presence of two null points, and hence two regions of low Alfv\'en speed, leads to the wave splitting. \cite{mclaughlin2006magnetohydrodynamics} then carried out a further investigation in a magnetic topology created by two magnetic dipoles, resulting in a more physical X-type null point setup by ensuring that the magnetic field strength becomes smaller at a large distance from the null point, which was not the case in \cite{mclaughlin2004mhd,mclaughlin2005mhd}. It was found that the central section of the wave exhibited the same wrapping behavior of the two previous investigations, but that the outer sections of the wave split away from this central accumulation, ultimately escaping the null point. 

Note that null points naturally break the low-$\beta$ condition in the corona, as magnetic pressure significantly drops at null points. \cite{mclaughlin2006mhd} removed the low plasma-$\beta$ approximation, finding that the fast magnetoacoustic wave is attracted towards the null point, generating a slow magnetoacoustic wave as it crosses the $\beta = 1$ layer. \cite{mclaughlin2006mhd} also found that the fast magnetoacoustic wave can pass through the null point in this setup, due to the nonzero sound speed. The present study considers a low-$\beta$ plasma, and as such, the equipartition layer is very close to the null point and mode conversion does not play a key role.

The propagation of MHD waves around null points has also been taken further than these linear, 2D systems: \cite{mclaughlin2009nonlinear} considered the effects of a nonzero plasma-$\beta$ when driving a nonlinear fast magnetoacoustic wave, finding that shock formation and collapse of the null point can drive oscillatory reconnection \citep{craig1991dynamic}. \cite{rickard1996current}, \cite{galsgaard2003numerical} and \cite{mclaughlin20083d} all investigated wave behavior in 3D. Readers are directed to \cite{mclaughlin2011mhd} for a review of MHD wave propagation around coronal null points without gravitational stratification.

MHD wave propagation around magnetic null points has also been investigated with gravitational stratification included. \cite{tarr2017magnetoacoustic} investigated the propagation of an initially nonlinear acoustic wave packet from the photosphere toward a magnetic null point with a magnetic dome topology in the lower corona, finding that a portion of the wave packet refracts toward the null point as a fast magnetoacoustic wave. \cite{pennicott2021conversion} considered MHD waves in a high-$\beta$, stratified atmosphere with magnetic null points, and investigated the behavior of shocks as they approach and pass through the $\beta = 1$ layer.

There are a number of other works combining the propagation of MHD waves around null points and gravitational stratification. \cite{santamaria2015magnetohydrodynamic} found that strong gradients in the Alfv\'en speed close to a null point can cause significant refraction, leading to downward-propagating waves in the corona. \cite{smirnova2016numerical} and \cite{gonzalez2022numerical} investigated how pressure pulses generated at the null point result in shocks and the formation of jets. \cite{yadav20223d} proposed a new method for MHD wave decomposition when considering the propagation of the MHD waves around coronal null points. \cite{liakh2025numerical} found that fast extreme-ultraviolet (EUV) waves generated by solar eruptions can perturb null points, resulting in magnetic reconnection.

\begin{table*}[t]
    \centering
    \begin{tabular}{|c|c|c|c|c|c|}
         \hline
         \textbf{Density Stratification} & \textbf{Numerical Resolution} & \textbf{$x$-domain} & \textbf{$y$-domain} & \textbf{Driver Position} & \textbf{Driver} \\
         \hline
         $C=0$ & $5000 \times 5000$ & $-10\leq x \leq 10$ & $-2\leq y \leq 18$ & Lower & Whole boundary \\ 
         $C=1$ & $5000 \times 2500$ & $-10\leq x \leq 10$ & $-2\leq y \leq 8$ & Lower & Spatially varying \\ 
         $C=1$ & $5000 \times 2500$ & $-10\leq x \leq 10$ & $-8\leq y \leq 2$ & Upper & Whole boundary \\ 
         $C=1$ & $2500 \times 2500$ & $-2\leq x \leq 8$ & $-10\leq y \leq 10$ & Left & Spatially varying \\ 
         \hline
    \end{tabular}
    \caption{The numerical resolution, location of boundaries, and driver descriptions of the simulations}
    \label{tab:Drivers}
\end{table*}

These works, along with previous studies without magnetic null points, have shown that the Alfv\'en speed profile plays a crucial role in the propagation of fast magnetoacoustic waves. \cite{1995SoPh..159..399N} found that fast magnetoacoustic waves are refracted toward regions of low Alfv\'en speed, and it is this behavior that leads to the results seen in \cite{mclaughlin2004mhd,mclaughlin2005mhd,mclaughlin2006magnetohydrodynamics}. \cite{brady2005damping} and \cite{verwichte2006seismology} also found that the fast magnetoacoustic wave could become trapped at local minima of the Alfv\'en speed caused by variations in density. 

\begin{figure}[t]
    \centering
    \includegraphics[width=0.9\linewidth]{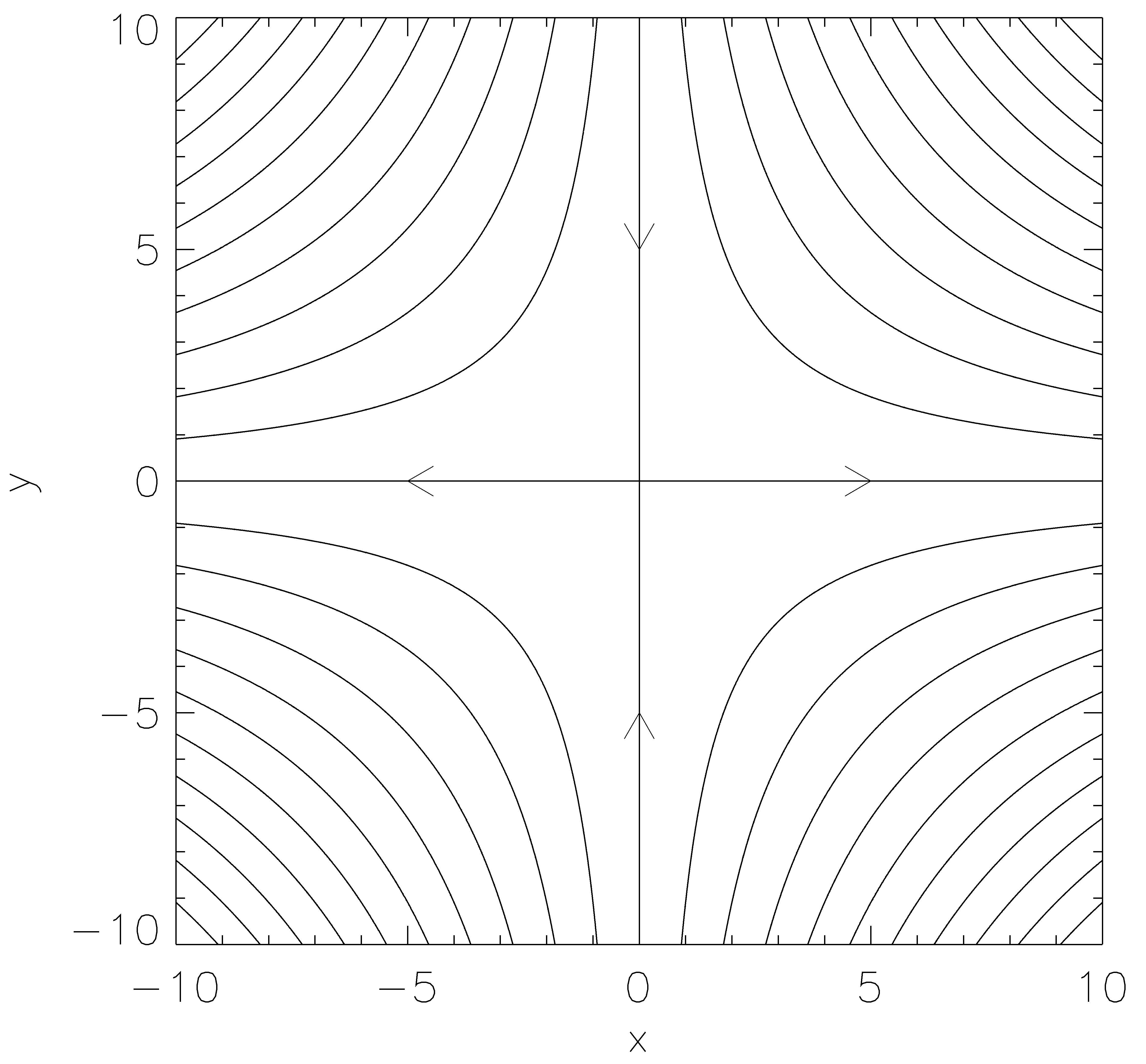}
    \caption{The equilibrium magnetic field, with arrows indicating the direction of the magnetic field along the separatrices.}
    \label{fig:1}
\end{figure}

This paper presents a linear, low-$\beta$ study of fast magnetoacoustic wave behavior within gravitationally stratified media, coupled with a simple, 2D X-type magnetic null point, which has not been reported upon previously. The manuscript is structured in the following manner: \S\ref{sec:method} describes the setup, the methodology, and an analysis of the Alfv\'en speed profile for a variety of levels of stratification; \S\ref{sec:lower} presents the results when the lower boundary is driven (both for the stratification-free and stratified cases), as well as an analysis of varying levels of stratification using the semianalytical Wentzel-Kramer-Brillouin (WKB) solution; \S\ref{sec:upper} and \S\ref{sec:left} present the results when the upper and left boundaries are driven, respectively, and discussion of results; and the conclusions are given in \S\ref{sec:conclusions}. Appendices \ref{sec:Appendix A} and \ref{sec:Appendix C} detail the derivation of the governing equation for the propagation of the linear fast magnetoacoustic wave while considering gravitational stratification. Appendix \ref{sec:Appendix B} demonstrates how these equations align with the literature. Simplifications that can be made to that equation and the derivation of the WKB solution can be found in \ref{sec:Appendix C}.

\section{Methodology} \label{sec:method}

This 2D numerical setup and relevant equations are set out in this section.

\subsection{Governing Equations}\label{subsec:Lare2D}

This paper considers the nondimensionalized, ideal MHD equations:

\begin{equation}
    \frac{D \rho}{D t} = -\rho \nabla \cdot \textbf{v},
    \label{eqn:1}
\end{equation}

\begin{equation}
    \frac{D\textbf{v}}{Dt} = \frac{1}{\rho}\left(\nabla \times \textbf{B} \right) \times \textbf{B} - \frac{1}{\rho}\nabla P + \textbf{g},
    \label{eqn:2}
\end{equation}

\begin{equation}
    \frac{D \textbf{B}}{D t} = \left( \textbf{B} \cdot \nabla \right)\textbf{v} - \textbf{B}\left( \nabla \cdot \textbf{v} \right),
    \label{eqn:3}
\end{equation}

\begin{equation}
    \frac{D \epsilon}{D t} = - \frac{P}{\rho} \nabla \cdot \textbf{v},
    \label{eqn:4}
\end{equation}
where $D/Dt$ is the advective derivative, $\rho$ is the plasma density, $\textbf{v}$ is the velocity, $\textbf{B}$ is the magnetic field, $P = \rho \epsilon \left( \gamma - 1 \right)$ is the plasma pressure, $\gamma$ is the ratio of specific heats, $\textbf{g}$ is gravitational acceleration, and $\epsilon$ is the specific internal energy.

An investigation will be conducted using both a numerical approach, using the Lare2D code \citep{arber2001staggered}, and a semianalytical approach, using the derivation that can be found in Appendices \ref{sec:Appendix A} and \ref{sec:Appendix C}.

Equations (\ref{eqn:1}) to (\ref{eqn:4}) are derived by nondimensionalizing the ideal MHD equations using the same nondimensionalization (though not the same notation) as in the LareXd manual \citep{arber2001staggered}. The following dimensionless quantities are defined: $x = \bar{L} x^*$, $\textbf{B} = \bar{B} \textbf{B}^*$ and $\rho =  \bar{\rho} \rho^*$. These nondimensionalizing constants are then used to define the nondimensionalization for the rest of the system, with $\bar{v} = \bar{B} /\sqrt{\mu_0\bar{\rho}}$, $\bar{P} = \bar{B}^2/\mu_0$, $\bar{t} = \bar{L}/\bar{v}$, $\bar{j} = \bar{B}/\mu_0\bar{L}$, $\bar{T} = \epsilon_0\bar{m}/k_B$, $\mu_{m0} = \bar{m}$ and $\epsilon_0 = \bar{v}^2$, such that $\textbf{v}=\bar{v}\textbf{v}^*$ and $t=\bar{t}t^*$.

\cite{2008SoPh..251..563M}

Numerical viscosity in the Lare2D code is controlled by two parameters, $\nu_1$ and $\nu_2$: $\nu_1$ controls linear viscosity, acting on all shocks, while $\nu_2$ acts on stronger shocks \citep{caramana1998formulations}. These parameters are problem dependent, and this work uses values of $\nu_1=0.1$ and $\nu_2=0.5$.

\subsection{Numerical Setup\label{subsec:setup}}
This study consists of four computational setups, with their main differences being the driven boundary and the location of the other three boundaries (see Table \ref{tab:Drivers}). The area of interest is close to the magnetic null point (which is found at the origin). This corresponds to an area given by $x, y \in \left[ -2,2\right]$, as was the case in \cite{mclaughlin2004mhd}. The location of the numerical boundaries is chosen such that the nondriven boundaries are sufficiently far away from the area of interest, to minimize the interference of boundary reflections. Note that the $y$-domain for the stratification-free case is twice the size as in the other three simulations, and this is to position the upper boundary even further away, preventing boundary reflections from propagating into the region of interest.

The equilibrium magnetic field is chosen to be a 2D simple X-type null point, as in \cite{mclaughlin2004mhd}, given by

\begin{equation}
    \textbf{B} = \frac{\bar{B}}{\bar{L}} \left(x,-y,0 \right),
    \label{eqn:5}
\end{equation}
where $\bar{B}$ and $\bar{L}$ represent characteristic field strength and length scale for magnetic field variations respectively. This can be seen in Figure \ref{fig:1}. It is important to note that, given that the magnetic field strength increases radially, this magnetic configuration becomes unphysical far away from the null point.

\begin{figure*}
    \centering
    \includegraphics[width=1\linewidth]{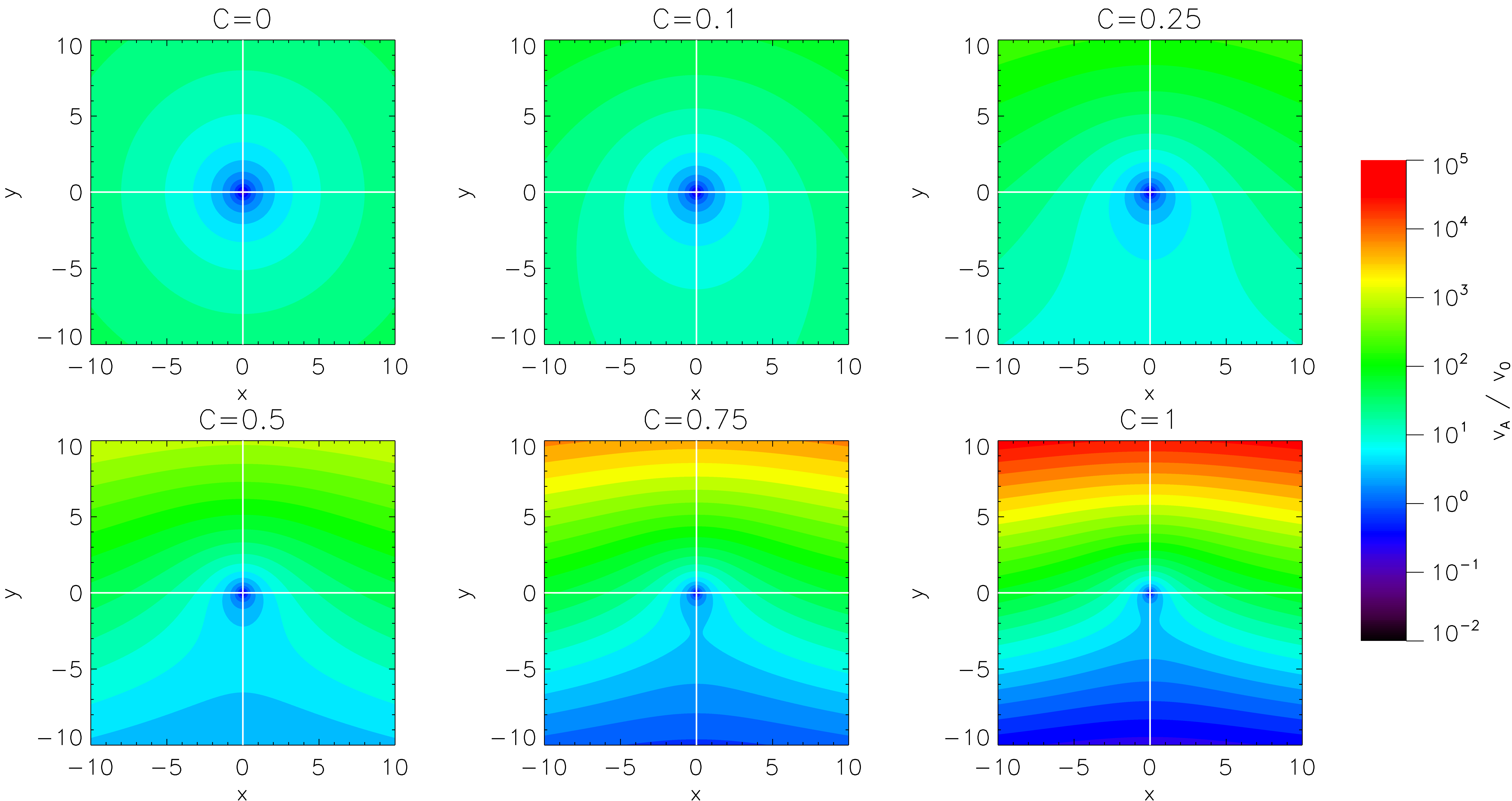}
    \caption{Contours of the Alfv\'en speed for a range of $C$ values, with a logarithmic color scale. Separatrices of the equilibrium magnetic field are overplotted in white. The magnetic null point is located at $\left[0,0\right]$.}
    \label{fig:2}
\end{figure*}

\begin{figure*}
    \centering
    \includegraphics[width=0.8\linewidth]{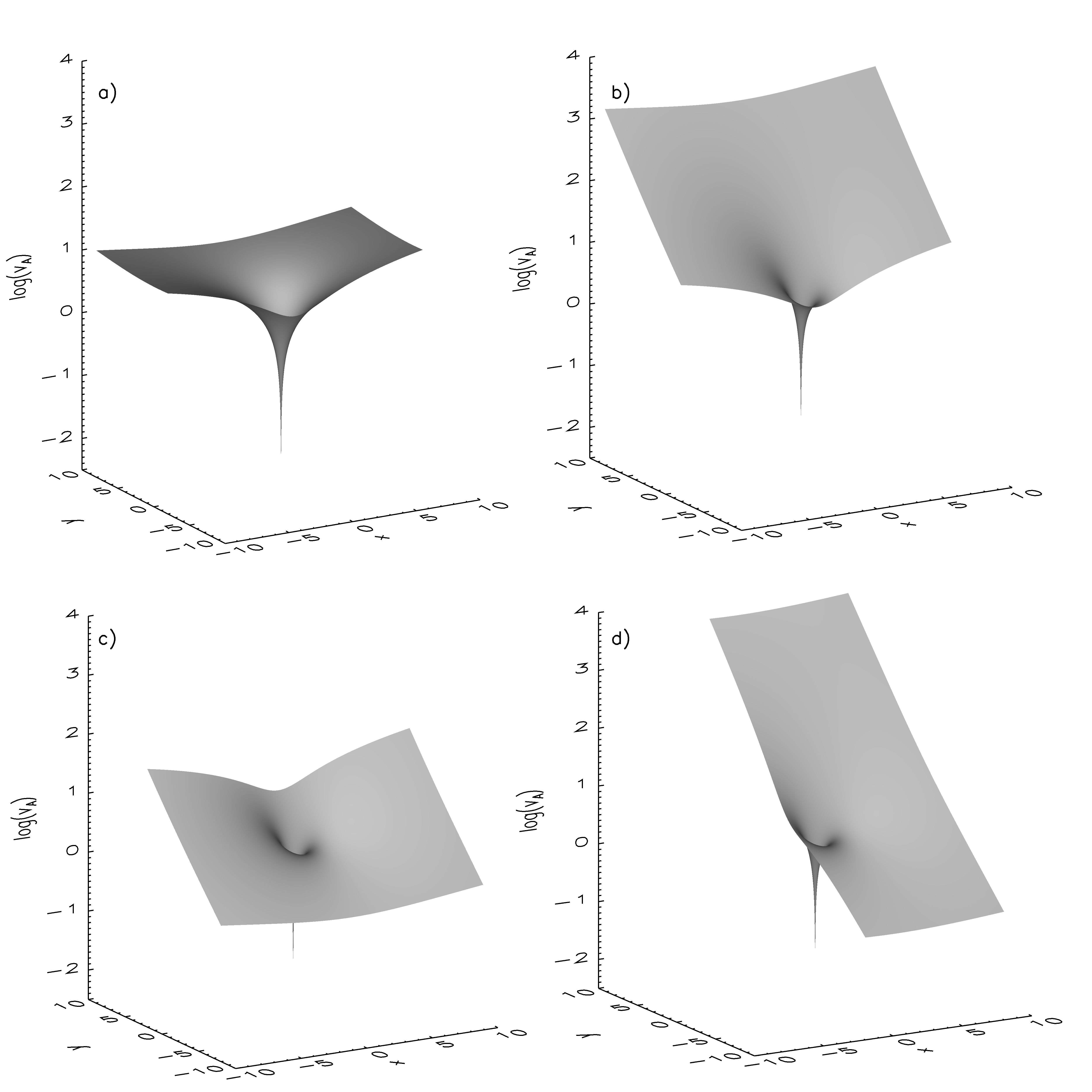}
    \caption{3D surfaces of $\text{log}\left(\text{v}_A\right)$, with a), b), c), and d) corresponding to the profiles in \S\ref{subsec:stratfree}, \S\ref{subsec:c1} (driving lower boundary, $C=1$), \S\ref{sec:upper} (driving upper boundary, $C=1$) and \S\ref{sec:left} (driving left boundary, $C=1$) respectively. Table \ref{tab:Drivers} details the domain boundaries. In each simulation, the magnetic null point is located at $\left[0,0 \right]$.}
    
    \label{fig:3}
\end{figure*}

A stratified density profile is derived by solving for pressure balance, giving

\begin{equation}
    \rho_0 = e^{-C\left(y+2\right)},
    \label{eqn:6}
\end{equation}
where $C=g/2T$, given that $\textbf{g} = \left(0,-g,0\right)$, and $T$ is temperature. The density profile is configured such that $\rho_0(y=-2)=1$, i.e. at the bottom of the $x, y \in \left[ -2,2\right]$ box. This stratified density profile is used in all three of the investigations. It is clear that when $C=0$, there is no stratification, and the system returns to the same setup as in \cite{mclaughlin2004mhd}.

Gravitational stratification introduces a finite density scale height to the system. This provides a measure of how rapidly the density $\rho_0$ changes with height $y$ \citep{roberts2019mhd}. In the setup investigated in this paper, the density scale height $H_0$ is given by $H_0 = 2T/g$, or $H_0 = 1/C$. This paper uses a value of $C=1$, giving a density scale height of $H_0 = 1$. This scale height corresponds to the density at the top of the area of interest (i.e. $y = 2$) being 4 orders of magnitude smaller than at the bottom of the box. It is clear from this definition that a large density scale height corresponds to a weakly stratified medium, while a small density scale height corresponds to a strongly stratified medium. 

\cite{mclaughlin2004mhd} found it helpful to consider $\text{v}_\perp$ and $\text{v}_\parallel$, where $\text{v}_\perp =\textbf{v} \times \textbf{B} \cdot \hat{\textbf{z}} = v_xB_y-v_yB_x$ and $\text{v}_\parallel = \textbf{v} \cdot \textbf{B} = v_xB_x+v_yB_y$ when investigating their system, and that approach is followed here. These parameters are related to the perpendicular and parallel components of velocity, respectively. 

Using this approach, Appendix \ref{sec:Appendix A} finds that the ideal, linearized MHD equations decouple into three equations: one for $v_z$, which corresponds to the Alfv\'en wave, given by Equation (\ref{eqn:A38}), and then a pair of equations that govern the (slow and fast) magnetoacoustic behavior, via one equation for $\text{v}_\perp$ given by Equation (\ref{eqn:A45}) and one for $\text{v}_\parallel$, given by Equation (\ref{eqn:A46}). Appendix \ref{sec:Appendix B} details how these two equations align with the literature, demonstrating that in both the horizontal and vertical magnetic field cases, the equations reduce to those of \cite{roberts2019mhd}. By driving $\text{v}_\perp$ and setting $\text{v}_\parallel(t=0)=0$, these equations allow the driving of a fast magnetoacoustic wave into the system (since the slow  magnetoacoustic wave cannot cross fieldlines, and thus cannot manifest in $\text{v}_\perp$). Note however that Equations (\ref{eqn:A45}) and (\ref{eqn:A46}) are coupled and so, without additional simplifications, driving $\text{v}_\perp$ will also generate a nonzero $\text{v}_\parallel(t)$, even if $\text{v}_\parallel(t=0)=0$ is set.

This work will focus on the propagation of the fast magnetoacoustic wave and so will focus on Equation (\ref{eqn:A45}), though simplifications to Equation (\ref{eqn:A45}) can be made, and are appropriate in the context of the solar corona. Those simplifications (details of which can be found in Appendix \ref{sec:Appendix C}) yield the following equation:

\begin{equation}
    \frac{\partial^2 \text{v}_\perp}{\partial t^2} = \frac{1}{\rho_0}\left(B_x^2+B_y^2\right)\left( \frac{\partial^2}{\partial x^2} + \frac{\partial^2}{\partial y^2}\right)\text{v}_\perp.
    \label{eqn:7}
\end{equation}
This is the same equation as Equation (\ref{eqn:C51})

Note that by assuming a constant density profile $\rho_0 = 1$, one arrives at the same wave equation as \cite{mclaughlin2004mhd} and \cite{mclaughlin2009nonlinear}.

Table \ref{tab:Drivers} details the configuration of the driver for each of the four simulations. A spatially varying driver is used for two of the simulations in order to prevent large gradients in the Alfv\'en speed being present during wave formation. Each  driver has the general form $\text{v}_\perp = A\text{sin}^2\left( \omega  t\right)$, with an amplitude of $A=0.001$ being chosen to ensure that the simulations primarily capture the propagation of a linear fast magnetoacoustic wave. The two drivers that include a spatially varying component are given by $\text{v}_\perp = A\text{sin}^2\left( \omega t\right)\text{sech}\left[\left(x/6 \right)^8\right]$, where the values of the denominator and exponent are chosen such that for $\left|x \right| \ge 7.5$, the amplitude of $\text{v}_\perp$ is less than $0.5\%$ that of the amplitude of $\text{v}_\perp$ for $-2\le x \le 2$. Here, $\omega=5\pi$ is chosen such that the waves are driven between $t=0$ and $t=0.2$, which results in a single positive pulse being driven into the system.

Zero gradient boundaries are employed for the nondriven velocity boundaries, as well as for all magnetic field components, temperature, energy, and the side boundaries for density. Given the exponential density profile, care must be taken to avoid any large discontinuities at the boundaries, and so second-order zero gradient boundary conditions in density ($\partial^2\rho/\partial y^2 = 0$) are implemented at the upper and lower boundaries, in order to minimize these effects.

\begin{figure*}[t]
    \centering
    \includegraphics[width=1\linewidth]{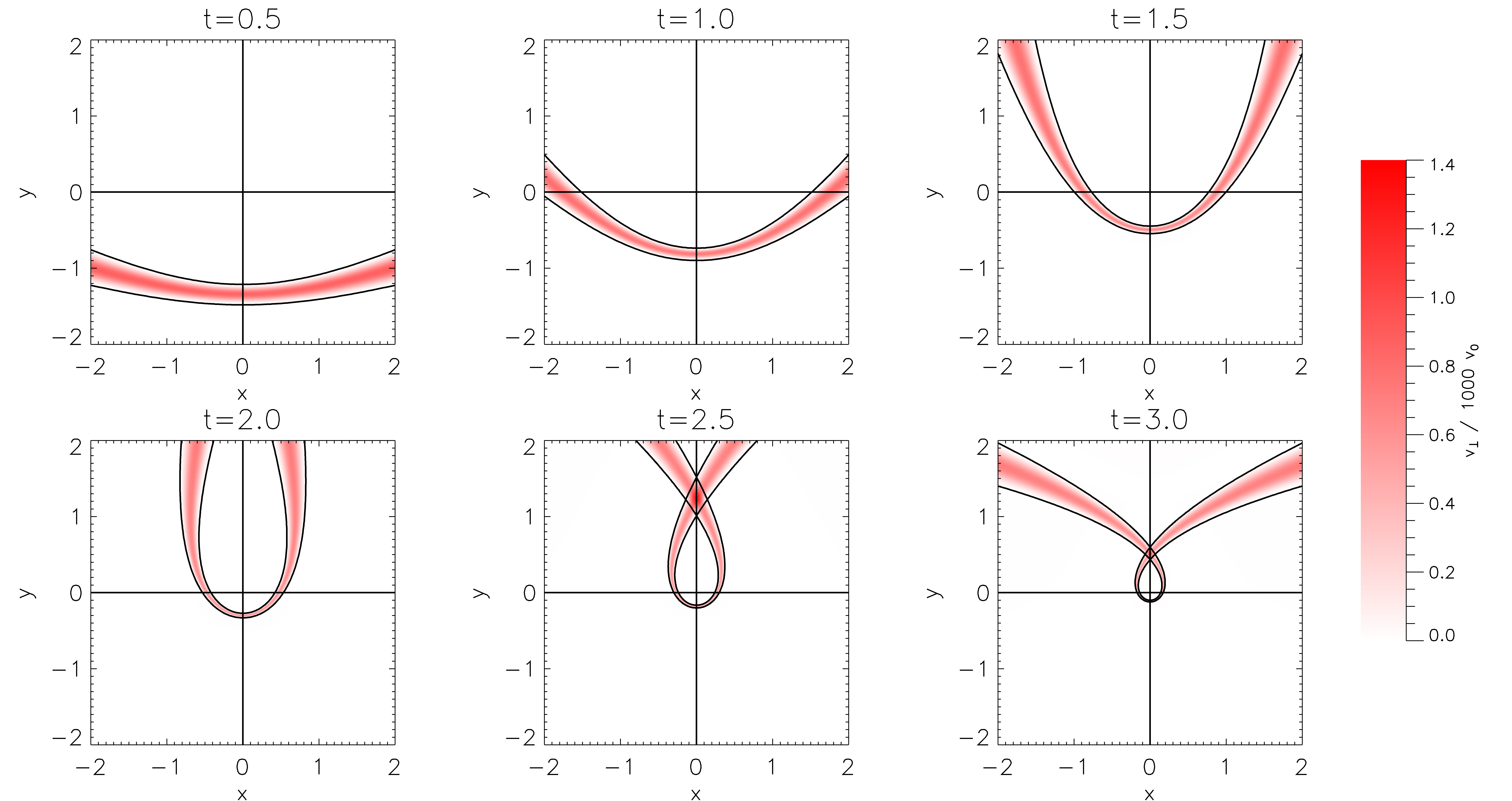}
    \caption{Contours of $\text{v}_\perp$ at six different times, with WKB solutions for the front and trailing edge of the wave overplotted as black lines, and separatrices of the equilibrium magnetic field are overplotted as straight black lines. In this case, the lower boundary is driven and there is no gravitational stratification.}
    \label{fig:4}
\end{figure*}

\subsection{Equilibrium Alfv\'en Speed Profiles\label{subsec:Alfven}}

Previous studies, such as \cite{1995SoPh..159..399N} and \cite{mclaughlin2006magnetohydrodynamics}, have found that the equilibrium Alfv\'en speed profile plays a crucial role in dictating the propagation of the fast magnetoacoustic wave. Variations in magnetic field strength and/or density cause variations in the Alfv\'en speed, given that $ \text{v}_A = \left|\textbf{B}\right|/\sqrt{\rho}$. Applying the equilibrium magnetic field and density profiles given in Equations (\ref{eqn:5}) and (\ref{eqn:6}) gives the following formula for the equilibrium Alfv\'en speed:

\begin{equation}
\text{v}_A = \sqrt{x^2+y^2} \cdot e^{{C}\left(y+2\right) / 2}.
\label{eqn:8}
\end{equation}
As such, a fast magnetoacoustic wave, which propagates at the Alfv\'en speed in a low-$\beta$ plasma, is directly affected by any variations in the magnetic field strength or density.

Figure \ref{fig:2} shows contours of the Alfv\'en speed profile for the equilibrium magnetic field, with a stratification-free case ($C=0$), and five cases consisting of varying degrees of stratification (i.e. varying $C$). The $C=0$ contour demonstrates the symmetry of the Alfv\'en speed profile, forming concentric circles around the null point, with the Alfv\'en speed increasing with increasing radius. The $C \neq 0$  cases demonstrate that with the introduction of gravitational stratification, the symmetry in $x$ remains present, but the symmetry in $y$ is broken. Increasing $C$ decreases the scale height, as $H_0 = 10$ when $C=0.1$, while $H_0 = 1$ when $C=1$. This decrease in scale height means that, for increasing $C$, the Alfv\'en speed at the top of the domain becomes multiple orders of magnitude greater than at the bottom of the domain. Consequently, driving waves from the lower, upper, and side boundaries of the stratified setup will lead to three different behaviors which are investigated in \S \ref{sec:lower}, \S \ref{sec:upper}, and \ref{sec:left}, respectively.

Figure \ref{fig:3} demonstrates the difference between the Alfv\'en speed profiles responsible for these three different behaviors (and how they all differ from the stratification-free case). It is clear from these profiles, as well as Figure \ref{fig:2}, that a wave driven from the lower boundary into a stratified medium is propagating against the Alfv\'en speed gradient, while a wave driven from the top boundary is propagating with the gradient, and a wave driven from the left boundary is propagating across it. Figures \ref{fig:2} and \ref{fig:3} also demonstrate the presence of a saddle point in the Alfv\'en speed, located at $\left[0,-2\right]$. In the case that the lower boundary is driven with gravitational stratification, this saddle point lies on the lower boundary (Table \ref{tab:Drivers}).

\begin{figure*}[t]
    \centering
    \includegraphics[width=1\linewidth]{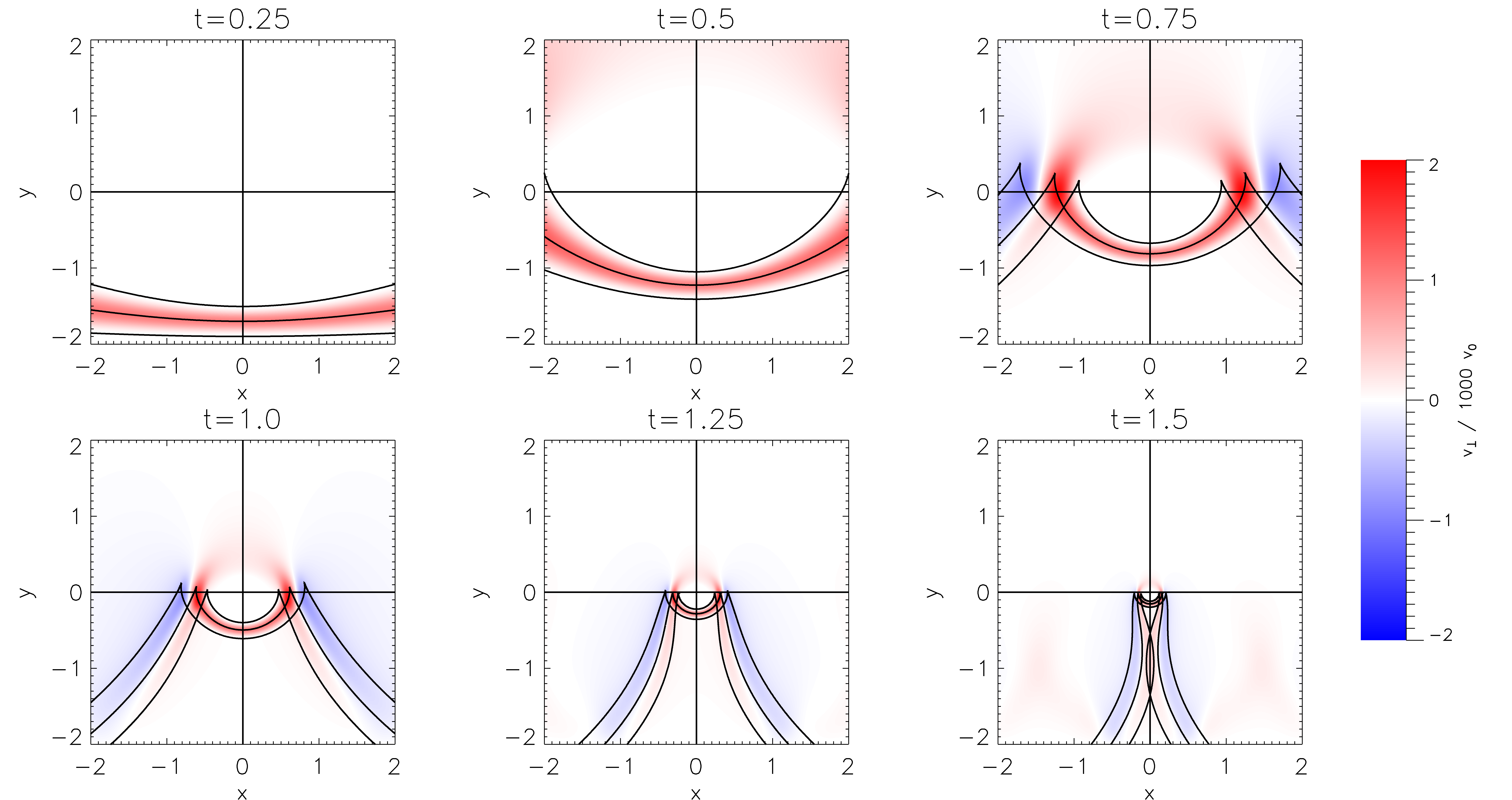}
    \caption{Contours of $\text{v}_\perp$ at six different times, with WKB solutions for the front, middle, and trailing edges of the wave overplotted as black lines, and separatrices of the equilibrium magnetic field are overplotted as straight black lines. In this case, the lower boundary is driven and there is gravitational stratification.}
    \label{fig:5}
\end{figure*}

\section{Results }\label{sec:lower}

Results from the numerical simulations and semianalytical WKB solution are presented in this section.

\subsection{Baseline case: No Stratification\label{subsec:stratfree}}

The propagation of the fast magnetoacoustic wave in the vicinity of the null point without stratification is considered, to establish a baseline for comparison. The numerical setup is described in \S\ref{subsec:setup} and Table \ref{tab:Drivers}. In addition to numerical simulations, one can also solve Equation (\ref{eqn:7}) semianalytically using the WKB approach and the full derivation of this can be found in Appendix \ref{sec:Appendix C}. Note that for the stratification-free case, only the lower boundary is driven, as driving any other boundary will give the same results due to the rotational symmetry in the system.

Figure \ref{fig:4} shows contours of $\text{v}_\perp$ at six different stages throughout the propagation of the wave, as well as the leading and trailing edges from the WKB solution overplotted. Note that there is agreement between the numerical simulation and the analytical WKB solution. The wave propagates toward the null point (seen at $t=0.5$), with the center of the wave propagating slower than the outer parts of the wave - which is a direct consequence of the Alfv\'en speed increasing with increasing $\left| x \right|$, as illustrated in Figures \ref{fig:2} and \ref{fig:3}. The wave continues to propagate upward (see $t=1.0$ and $t=1.5$), with the outer sections beginning to wrap around the null point (at $t=2.0$), which is a consequence of the wave refracting away from regions of higher Alfv\'en speed and toward regions of lower Alfv\'en speed.  

The wrapping continues, until (at $t=2.5$) the two outer sections of the wave cross one another, though due to the choice of $A=0.001$, no nonlinear effects occur. These outer parts then begin the same wrapping process (seen at $t=3.0$), but this time tighter to the null and propagating in the opposite direction, and if the simulation were run for longer (and ran in a domain that does not cut off at $y=-2$), a similar crossover event and change of direction would be observed, on a similar time frame to the crossover shown in Figure \ref{fig:4}. While the wrapping is happening, the central section of the wave continues to propagate toward the null point, in essence further tightening the wrapping effect.

It is important to note that as the wave gets closer and closer to the null point, it slows down. At the null point itself the Alfv\'en speed is zero, and as such, the wave never actually reaches the null point (for the linear case).

\subsection{Driving Lower Boundary: Stratified Case $(C=1)$}\label{subsec:c1}

\begin{figure*}
    \centering
    \includegraphics[width=1\linewidth]{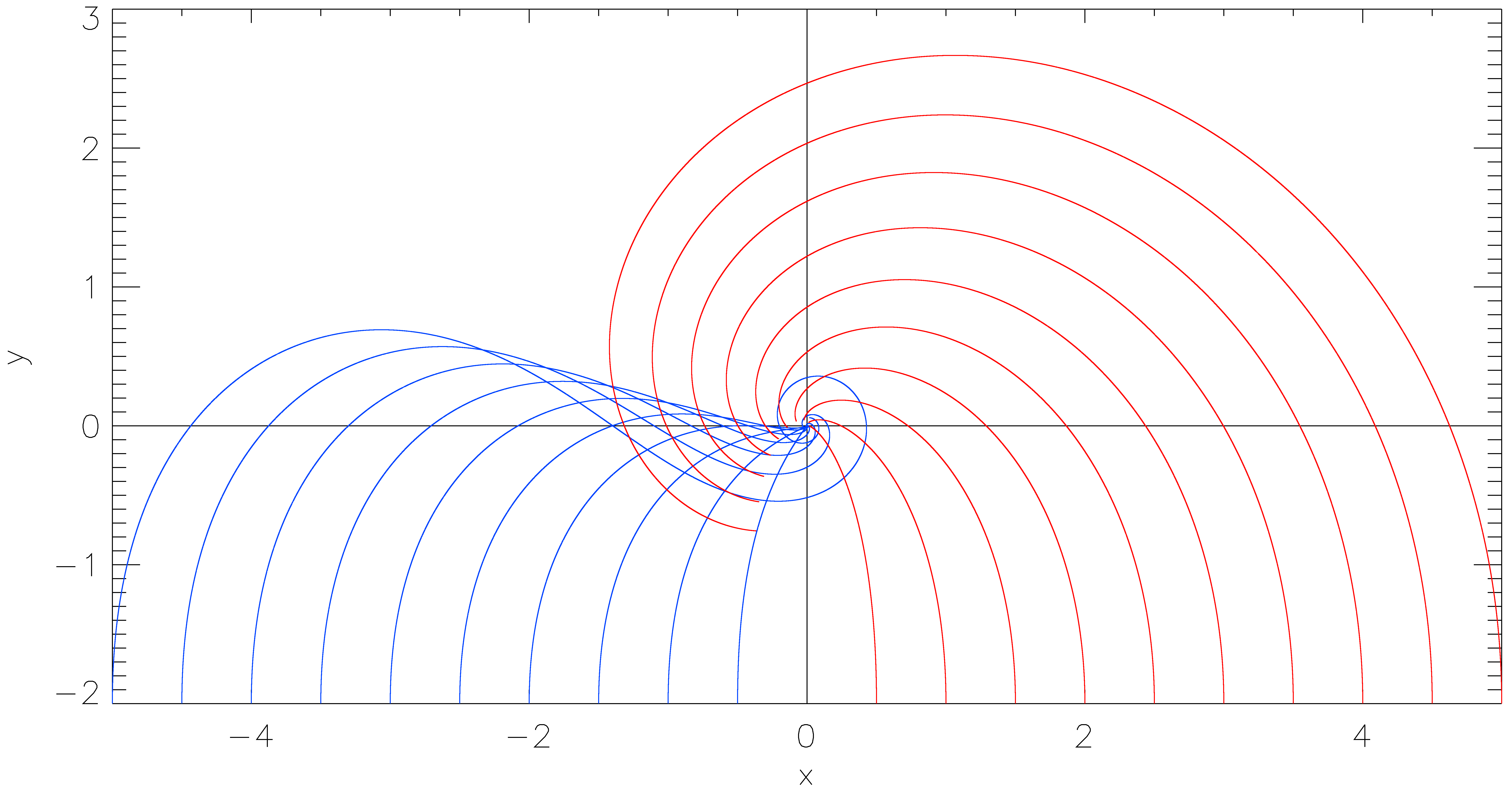}
    \caption{WKB ray paths for a range of starting points for both the $C=1$ stratified (blue) and nonstratified (red) cases, with the equilibrium magnetic field separatrices plotted in black.}
    \label{fig:6}
\end{figure*}

This section considers the propagation of the fast magnetoacoustic wave in the vicinity of the null point with stratification. The numerical setup is described in \S\ref{subsec:setup} and Table \ref{tab:Drivers}, with part of the lower boundary being driven. The density profile given in Equation (\ref{eqn:6}) is used, with $C=1$. This yields a density scale height of $1$, which corresponds to steep density and Alfv\'en speed gradients in the system, as shown in Figures \ref{fig:2} and \ref{fig:3}. Thus, at the lower edge of the area of interest $\rho_0(x,y=-2) = 1$, while at the upper edge, $\rho_0(x,y=2) = e^{-4}$.

Figure \ref{fig:5} shows contours of $\text{v}_\perp$ at six different stages throughout the wave's propagation, as well as the leading, middle, and trailing edges from the WKB solution overplotted. At $t=0.5$, the wave has reached closer to the null point than was the case in the stratification-free case shown in Figure \ref{fig:4}. This is a direct consequence of the wave propagating through areas of higher Alfv\'en speed than were encountered without stratification. However, from $t=0.75$ onward, the behavior is fundamentally different to the $C=0$ case. The wavefront is no longer a smooth curve, but instead a piecewise smooth curve. The outer parts of the wave, which move toward one another and cross at $t=1.5$ (though due to the linear nature of the wave, no nonlinear effects occur), are comprised of a section of positive $\text{v}_\perp$ and a section of negative $\text{v}_\perp$. At the same time, the cusps in the wavefront continue to move toward the null point, similar to the wrapping effect seen in Figure \ref{fig:4}. The WKB solution exhibits the same piecewise smooth wavefront, capturing the behavior of the sections of positive and negative $\text{v}_\perp$. The formation of the piece-wise smooth wavefront is a consequence of the refraction effect, leading to the crossing over of the WKB ray paths, which is explored further in \S\ref{subsec:WKBcrange}. 

There is good agreement between the numerical simulations and the analytical WKB solution, capturing the behavior of the central section of the wave, as well as capturing the piecewise smooth nature. It is, however, clear that the WKB solution does not fully align with the contours, particularly when it comes to the region of behavior above $x=0$. This is likely a consequence of the spatially varying driver, and the spreading out of the wave, a phenomenon which the WKB solution cannot capture. 

\subsection{Caustics and a Parameter Study of $C$ using the WKB Approximation}\label{subsec:WKBcrange}

Having looked at the propagation of the fast magnetoacoustic wave in the vicinity of a magnetic null point with and without gravitational stratification, the next step is to use the semianalytical WKB solution to compare the two cases and understand the formation of the piecewise smooth wavefronts seen in Figure \ref{fig:5}.

Figure \ref{fig:6} combines ray paths from both the $C=1$ stratified (blue), and stratification-free (red) cases, and uses the symmetry of the system in the $y$-axis to allow for direct comparison. In both cases, the ray paths are initialized from $y=-2$ equidistantly along $x$, and for the stratification-free case (red), it is seen that whilst the space between each ray path reduces, this is a gradual process, becoming more noticeable toward the end of the ray paths, as they approach the null point. Crucially, these (red) ray paths never cross one another.

Ray tracing follows the time evolution of waves, resulting in the path length at a given time instance. This allows for the determination of wave front surfaces. Fast magnetoacoustic waves refract away from areas with a higher Alfv\'en velocity \citep{1995SoPh..159..399N}. As a result, neighboring rays sample the equilibrium Alfv\'en speed profile at different rates, leading to variation in their refraction. Consequently they cross each other, at points called caustics. In a system containing several caustics, the caustics can be connected by a caustic curve \citep{chambersdictionary1999}. These caustics and caustic curves can result in cusps that change the wave front from a smooth curve into a piecewise smooth curve \citep{stamnes1991fast}. The blue ray paths in Figure \ref{fig:6}, corresponding to the $C=1$  stratified case, begin to refract away from the region of increasing equilibrium Alfv\'en speed seen in Figures \ref{fig:2} and \ref{fig:3}, and this leads to caustics being present. The caustics become more prevalent closer to the null point.

\begin{figure*}
    \centering
    \includegraphics[width=1\linewidth]{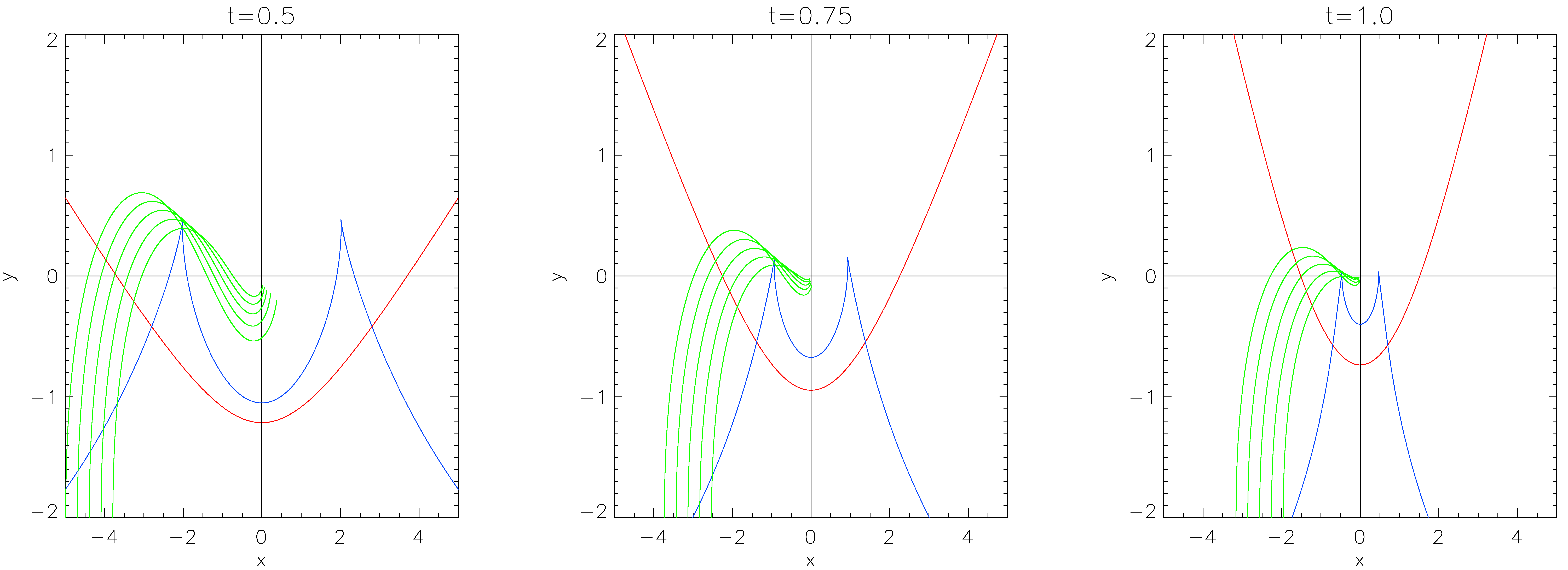}
    \caption{WKB wavefronts at three different times, for both the $C=1$ stratified (blue) and nonstratified (red) cases, with WKB ray paths (green) demonstrating that the positions of the cusps in the wavefronts correspond to the crossover points of the ray paths. The equilibrium magnetic field separatrices are plotted as black lines.}
    \label{fig:7}
\end{figure*}

Figure \ref{fig:7} demonstrates the link between the caustic points and the location of the cusp in the wavefront. The green ray paths all move upward after initialization, before reaching a turning point and starting to move downward, after which they start to wrap around the null point. The downward movement is a manifestation of the refraction away from the regions of increasing background Alfv\'en speed that are being propagated into. The further left along the $x$-axis the starting part of the ray path is, the higher along $y$ that turning point will be, due to the larger magnetic field in the numerator of the Alfv\'en speed. Despite the ray paths all following a smooth behavior, the wavefront is piecewise smooth, with two (symmetrically located about $x=0$) cusps forming as a result of the caustics in the ray paths.

\begin{figure*}
    \centering
    \includegraphics[width=1\linewidth]{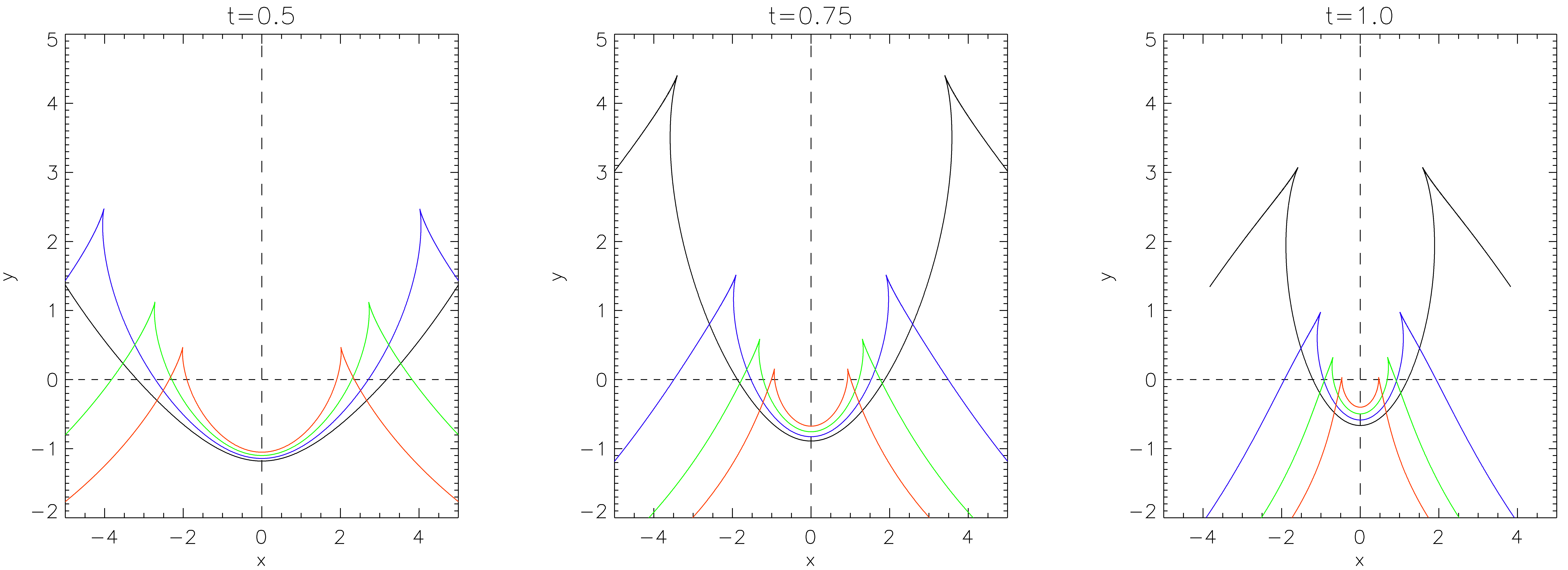}
    \caption{WKB wavefronts for $C=1/4, H_0=4$ (black); $C=1/2, H_0=2$ (blue); $C=3/4, H_0=4/3$ (green); and $C=1, H_0=1$ (red), demonstrating that for increasing C, the position of the cusp at a given time is closer to the null in both $x$ and $y$. The equilibrium magnetic field separatrices are plotted as dotted black lines.}
    \label{fig:8}
\end{figure*}

Figure \ref{fig:8} compares wavefronts for varying values of $C$ at three fixed times, to analyze how changing the value of C affects the location of the cusps in the wavefront. For a higher value of $C$, the location of the cusp has lower $\left| x \right|$, and lower $y$, i.e. the higher the value of $C$, the closer the cusps are to the null point. As $C$ increases, the density scale height $H_0$ decreases, while the magnetic field strength stays the same. The decreasing scale height causes larger gradients in Alfv\'en speed, and these gradients affect refraction.

\subsection{Driving Upper Boundary $(C=1)$}\label{sec:upper}

Introducing gravitational stratification removes the vertical symmetry from the system, and as such, this section now drives the upper boundary of the area of interest. The numerical setup follows that described in \S\ref{subsec:setup} and Table \ref{tab:Drivers}, with $C=1$ and the entire top boundary being driven.

Figure \ref{fig:9} shows contours of $\text{v}_\perp$ at six different stages throughout the propagation of the wave, as well as the leading and trailing edges from the WKB solution overplotted. Figures \ref{fig:2} and \ref{fig:3} demonstrate the Alfv\'en speed profile of the system, highlighting the clear difference between driving a wave from the lower boundary, as has been done previously, and the upper boundary. The wave initializes at a much higher Alfv\'en speed, and this is clear from the fact that the outer sections of the wave in the $t=0.25$ contour from Figure \ref{fig:9} have already begun to cross the $y=0$ line, which took longer when driving the lower boundary. As a result of this more rapid propagation, the signature wrapping effect begins to occur sooner than in \S\ref{subsec:stratfree}. Not only does the whole wave, especially the outer sections, propagate more rapidly than has been previously seen, but the behavior is also noticeably different. The presence of the saddle point at $[0,-2]$, discussed in \S\ref{subsec:Alfven}, affects the propagation of the wave (seen from $t=1$ onward), causing the part of the wave close to it to slow down (an effect similar to that of the null point). The outer sections of the wavefront cross one another at $t=1.25$ and, similar to Figure \ref{fig:4}, no nonlinear effects are generated. Note that the wrapping around the null point is a lot tighter than has been previously seen, and again, this is a direct consequence of the wave driven from the top boundary being initialized into a region of greater Alfv\'en speed than the wave driven from the lower boundary (see Figures \ref{fig:2} and \ref{fig:3}).

\begin{figure}
    \centering
    \includegraphics[width=0.9\linewidth]{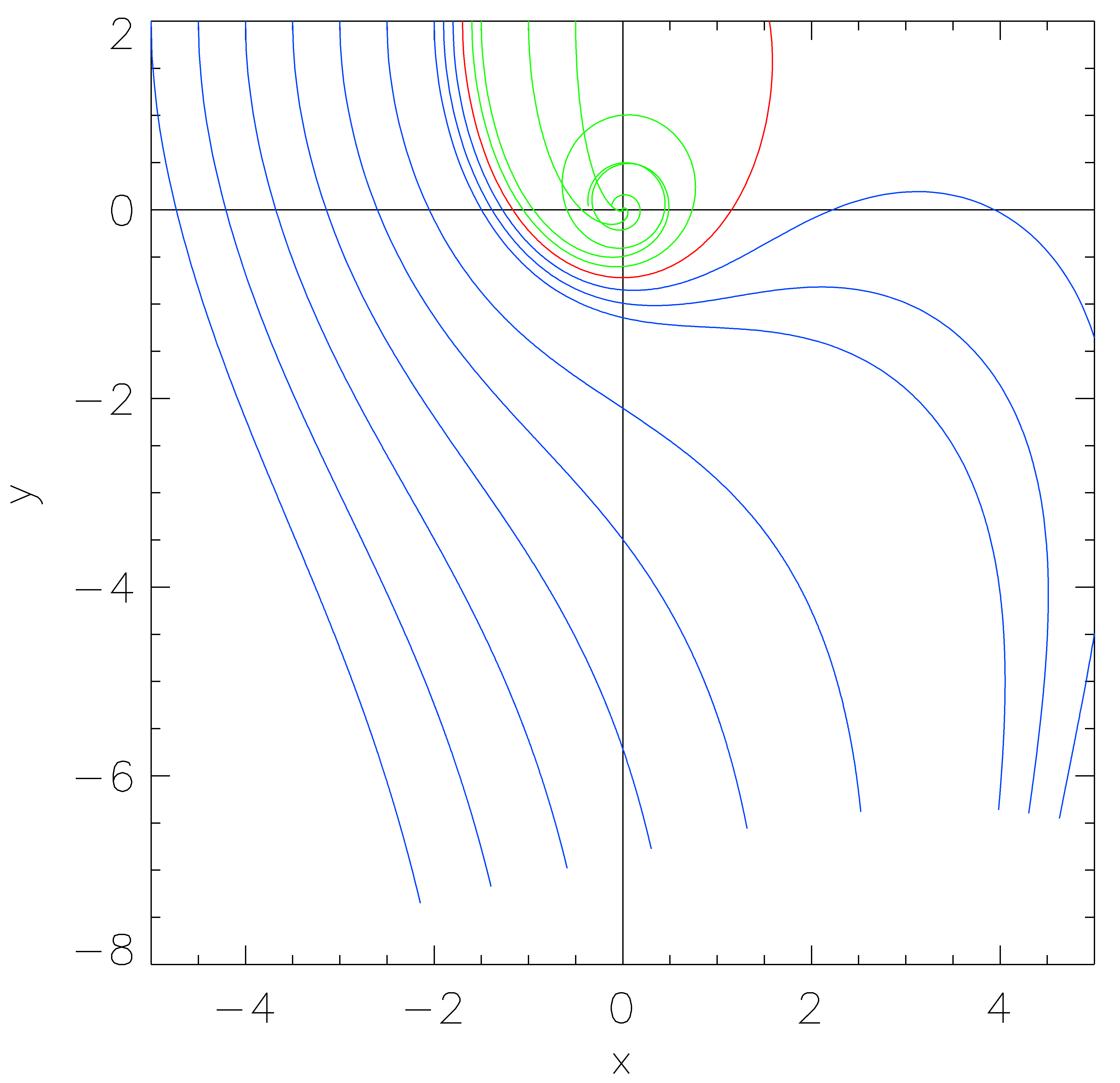}
    \caption{WKB ray paths for a range of starting points within the numerical domain when $C=1$ and the upper boundary is driven. There are three distinct behaviors, indicated by color, with equilibrium magnetic field separatrices in black. Blue ray paths (generated from $x<-1.7, y=2$) are influenced by the null point but escape, green ray paths (generated from $|x|<1.7, y=2$) become trapped by the null point, and the red ray path (generated from $x=-1.7, y=2$) orbits the null point.} 
    \label{fig:10}
\end{figure}

Figure \ref{fig:10} shows ray paths for a variety of starting points along $y=2$ within the numerical domain, and is split into three different colors, corresponding to different behaviors. Blue ray paths (generated for $x<-1.7, y=2$) begin to refract toward the null point ($x=0,y=0$), but are also influenced by the saddle point ($x=0,y=-2$). This behavior leads to ray paths that look as though they are going to wrap around the null, but then leave its influence and begin to refract toward the saddle point. Green ray paths (generated for $|x|<1.7, y=2$) exhibit an entirely null-point-dominated behavior, propagating quickly in the downward direction before refracting toward the regions of lower Alfv\'en speed close to the null point. The red ray path, with a starting point of $x=-1.7, y=2$, combines the two behaviors. This acts as a dividing line, where any ray paths starting from $x < -1.7$ do not wrap around the null point, similar to the blue ray paths, whilst any ray paths starting from $x > -1.7$ eventually wrap around the null point, the same behavior shown by the green ray paths. The blue ray paths in Figure \ref{fig:10} are all of different lengths for the same elapsed time. This highlights the different Alfv\'en speeds experienced along each ray path.

\subsection{Driving Left Boundary $(C=1)$}\label{sec:left}

This section considers the propagation of the fast magnetoacoustic wave in the vicinity of the null point, but this time driving the left boundary. The numerical setup follows that described in \S\ref{subsec:setup} and Table \ref{tab:Drivers}, with $C=1$ and the spatially varying driver.

\begin{figure*}[t]
    \centering
    \includegraphics[width=1\linewidth]{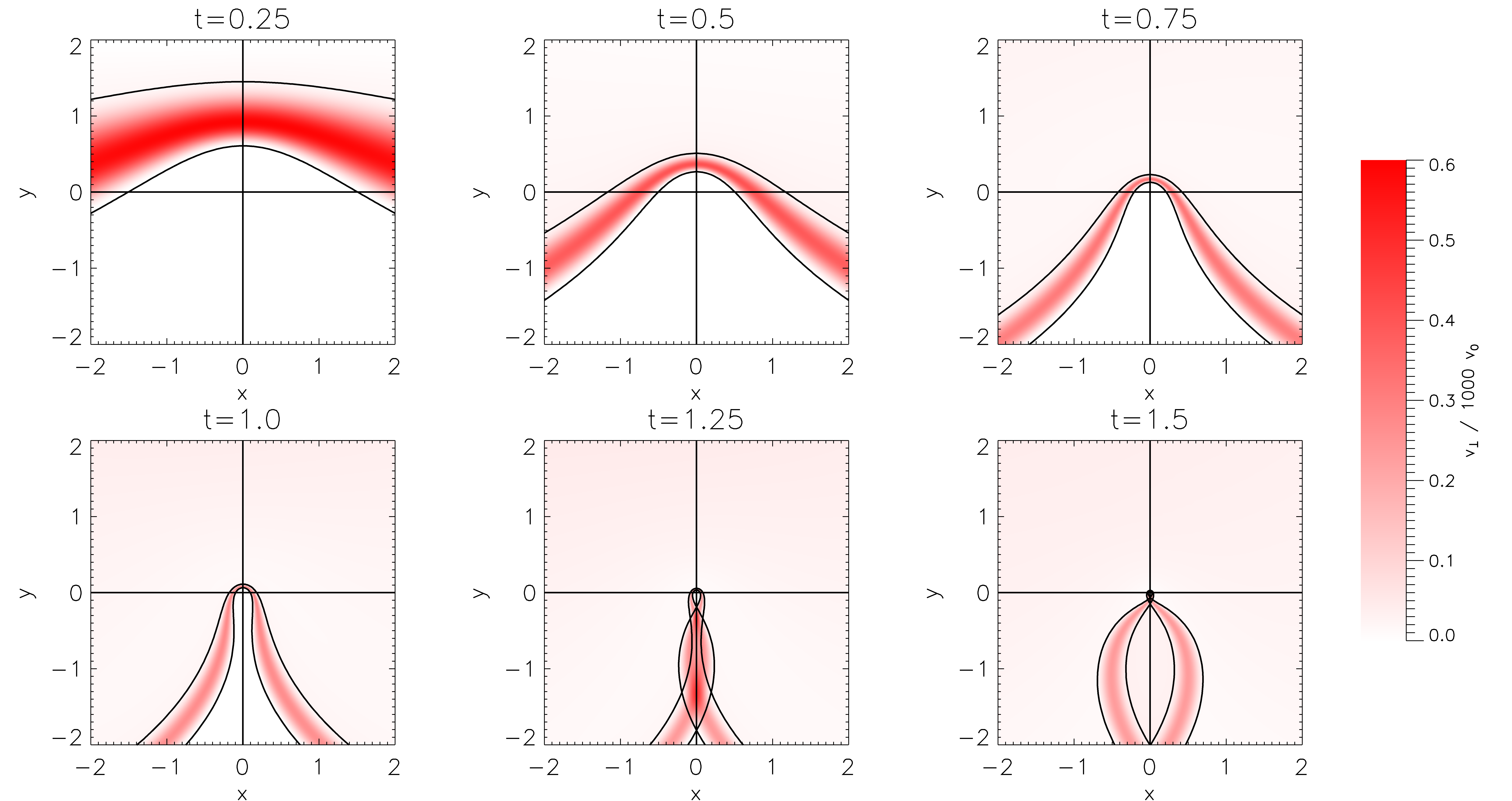}
    \caption{Contours of $\text{v}_\perp$ at six different times, with WKB solutions for the front and trailing edge of the wave overplotted as black lines, and separatrices of the equilibrium magnetic field are overplotted as straight black lines. In this case, the upper boundary is driven and there is $C=1$ gravitational stratification.}
    \label{fig:9}
\end{figure*}

\begin{figure*}
    \centering
    \includegraphics[width=1\linewidth]{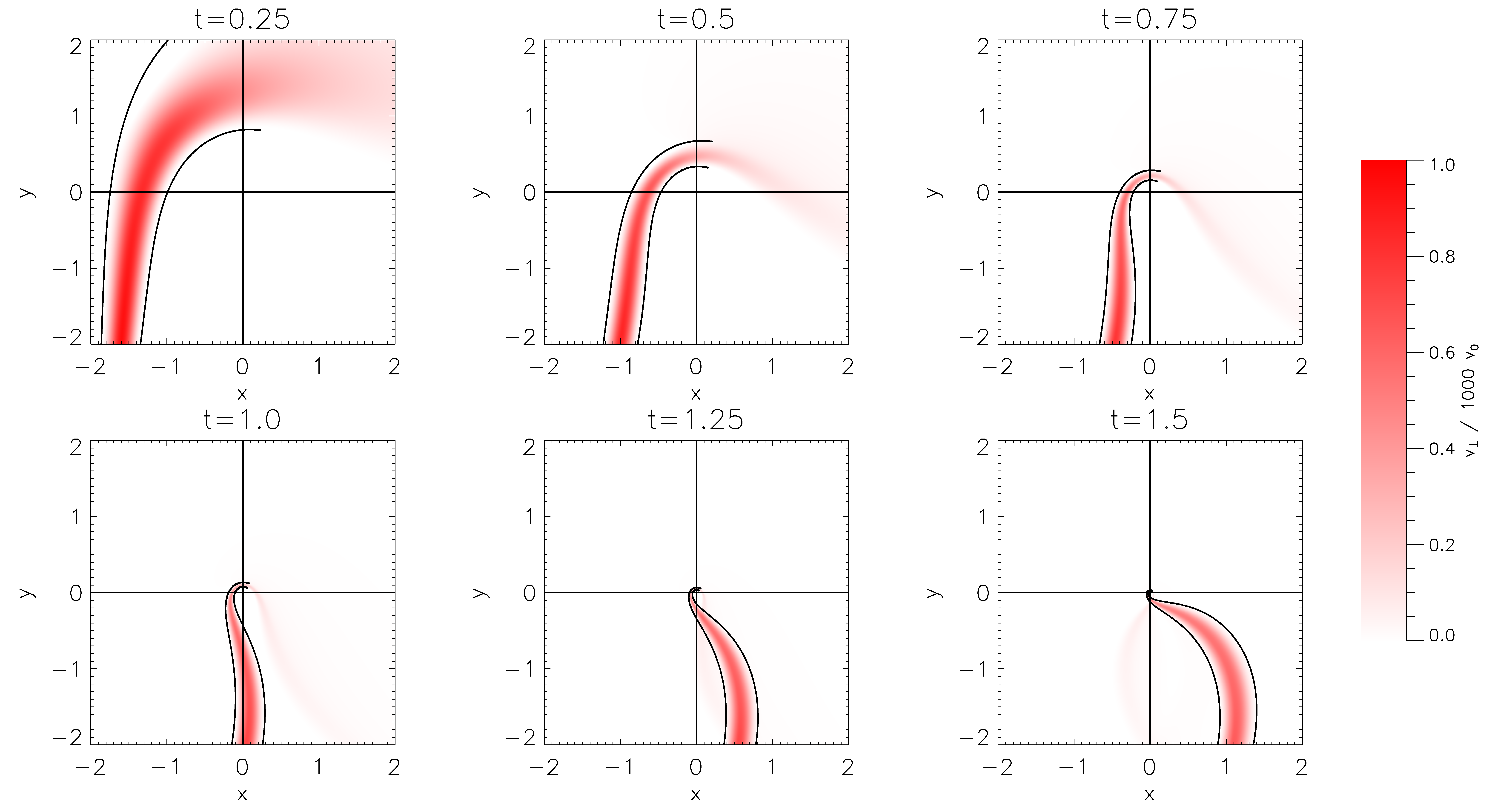}
    \caption{Contours of $\text{v}_\perp$ at six different times, with WKB solutions for the front and trailing edges of the wave overplotted as black lines, and separatrices of the equilibrium magnetic field are overplotted as straight black lines. In this case, the left boundary is driven and there is $C=1$ gravitational stratification.}
    \label{fig:11}
\end{figure*}

Figure \ref{fig:11} shows contours of $\text{v}_\perp$ at six different stages throughout the wave's propagation, as well as the leading and trailing edges of the wavefront obtained from the WKB solution overplotted. Note that the WKB solution in this section was only initialized between $y=-8$ and $y=8$ (i.e. the spatial domain of the driven wave) as the large Alfv\'en speeds upward of $y=8$ require an unfeasibly small $\Delta t$ in order to avoid numerical errors.

Figure \ref{fig:11} shows that, in the early stages, the upper section of the wave propagates significantly faster than the lower section - a direct consequence of the Alfv\'en speed gradient seen in the Alfv\'en speed profiles in Figures \ref{fig:2} and \ref{fig:3}. The faster upper part of the wave begins to refract away from the areas of high Alfv\'en speed immediately, and the wrapping effect begins to occur (see the first two panels). Meanwhile, the slower lower part continues to propagate rightward (taking all six panels to do so), and is influenced by both the null point and the Alfv\'en speed saddle point, with the sections of the wave close to either of these points propagating much slower than the center of the lower section. This leads to a similar behavior to that observed in \S\ref{sec:upper}. As expected due to the speed of propagation, the wrapping effect around the null point is much tighter than was observed in the stratification-free case. Figure \ref{fig:11} also shows that the wave has spread out a bit, seen as a faint red wavefront, and this is again a consequence of the spatially varying driver. Figure \ref{fig:11} demonstrates behavior comparable with Figure \ref{fig:9}, lacking the cusps seen in Figure \ref{fig:5}. It is clear from Figure \ref{fig:11} that agreement between the numerical simulations and the semianalytical WKB solution continues to be good, though the range that the WKB solution is ran over could be increased (with a sufficiently small $\delta t$) to capture this.

\begin{figure}
    \centering
    \includegraphics[width=0.9\linewidth]{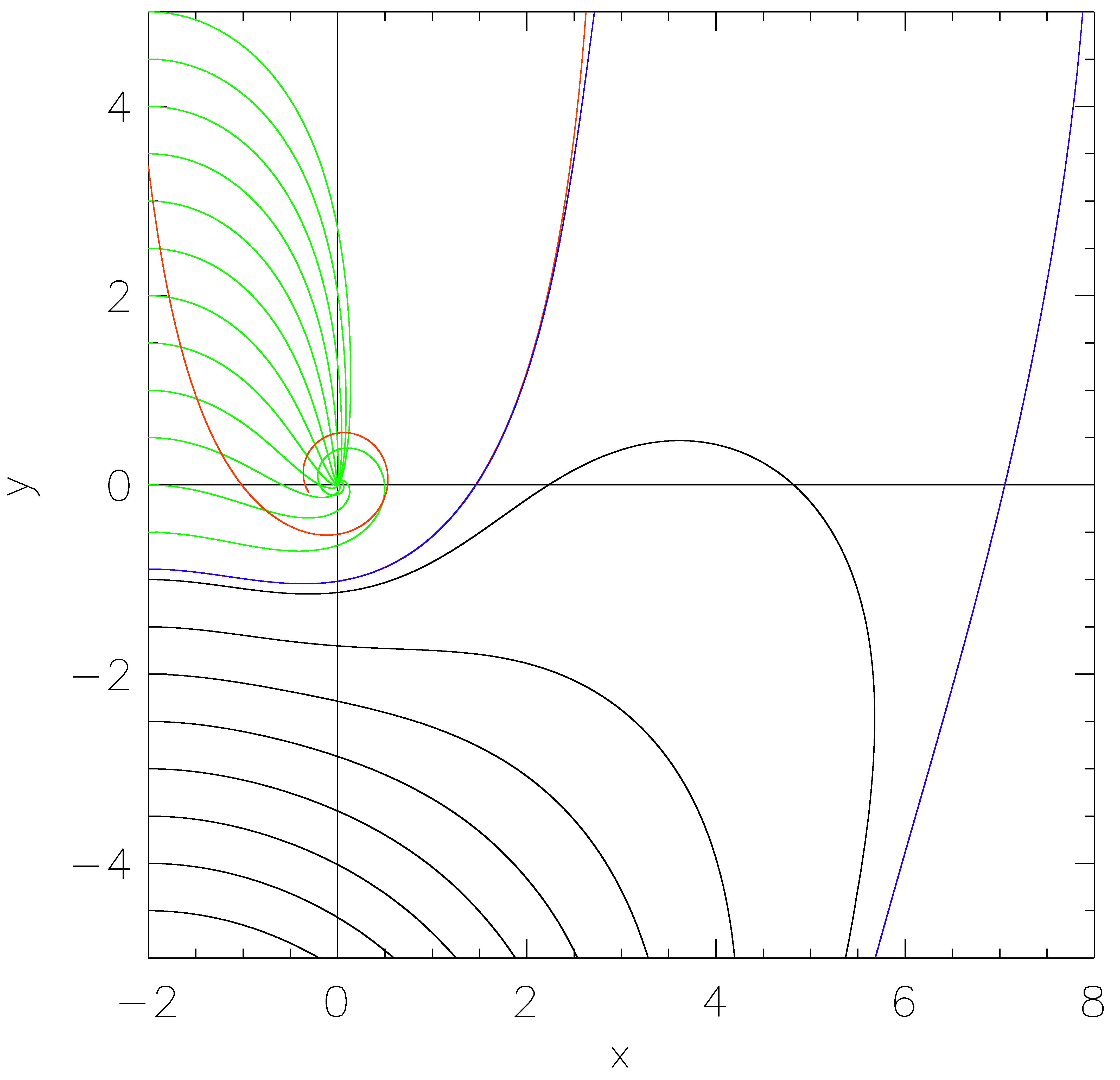}
    \caption{WKB ray paths for a range of starting points within the numerical domain when $C=1$ and the left boundary is driven. There are two distinct behaviors: black ray paths (generated for $x=-2, y<-0.8888$) are influenced by the null point but escape while green ray paths (generated for $x=-2, y>-0.8892$) become trapped by the null point. The red ($x=-2, y=-0.8888$) and blue ($x=-2, y=-0.8892$) ray paths indicate the crossover between the two behaviors. The equilibrium magnetic field separatrices are plotted in black.}
    \label{fig:12}
\end{figure}

Figure \ref{fig:12} shows ray paths for a variety of starting points within the numerical domain, and is split into four different colors, corresponding to different behaviors. Black ray paths are influenced somewhat by the null point, before  refracting away from the areas of high Alfv\'en speed far enough either side of the null point, and falling away. Green ray paths are initialized into a region of high Alfv\'en speed, and as such, refract away almost instantly, hence seeming to fall toward the null point (with the exception of the ray paths originating sufficiently close to the $y=0$ line, which seem to move downward at first before refracting away from the region of increasing Alfv\'en speed and toward the null point). The blue and red ray paths represent the divide between the two behaviors. The red ray path (with starting point $y=-0.8888$) propagates rightward and then upward, before refracting away from the ever increasing Alfv\'en speed and eventually falling back toward the null point, in a similar behavior to the ray paths in green. The blue ray path (with starting point $y=-0.8892$) follows a near identical path to the red ray path initially, but refracts rightward and ends up falling away in a similar way to the ray paths in black. 

\section{Discussion and Conclusions} \label{sec:conclusions}

This paper investigates the behavior of the linear fast magnetoacoustic wave around a magnetic null point with gravitational stratification. The ideal MHD equations are solved numerically using the Lare2D code \citep{arber2001staggered}. The magnetic topology considered was first investigated in \cite{mclaughlin2004mhd} without gravitational stratification. Results from the stratification-free case ($C=0, H_0 \rightarrow \infty$) are included here to form a comparison with the results when stratification is considered ($C \neq 0$, finite $H_0$).

To compare with the numerical solutions, the  ideal MHD equations are also solved semianalytically, using the WKB solution. A pair of wave equations, namely Equations \ref{eqn:A45} and \ref{eqn:A46} are derived, for $\text{v}_\bot$ and $\text{v}_\parallel$. These equations are proven to reduce to equations from existing literature \citep{roberts2019mhd} in Appendix \ref{sec:Appendix B}. This paper focuses on the propagation of the fast magnetoacoustic wave, and as such focuses on Equation \ref{eqn:A45}. Simplifications, found in Appendix \ref{sec:Appendix C}, detail how this equation reduces to Equation \ref{eqn:7}, and how this equation results in a semianalytical WKB solution.

Additionally, the WKB solution was used to investigate and explain new behavior seen in the numerical simulations. The WKB solution can be utilized in two different ways: fixing the starting point and considering all values of $s$ (the parameter relating to distance along the characteristic curve) to generate a WKB ray path, whereas fixing the value of $s$ and considering all starting points generates a WKB wavefront. In ray tracing terminology, the point at which two ray paths cross is called a caustic. Examples of caustics can be seen in Figures \ref{fig:6} and \ref{fig:7}. In a system with many caustics, the curve connecting the caustics is called a caustic curve. The effects of these caustics and caustic curves can be seen in the wavefront, with the formation of cusps, which change the wavefront from a smooth curve to a piecewise smooth curve. Cusps in wavefronts are present in Figures \ref{fig:5}, \ref{fig:7} and \ref{fig:8}.

When considering the problem without gravitational stratification, the background or equilibrium Alfv\'en speed is entirely dependent on the magnetic field (due to the constant density), and for the case of the simple X-point, this means that the Alfv\'en speed profile is radially symmetric, which is demonstrated in Figure \ref{fig:2}. This radial symmetry, combined with the fact that fast magnetoacoustic waves are refracted toward regions of low Alfv\'en speed \citep{1995SoPh..159..399N}, leads to the behavior seen in Figure \ref{fig:4}. The outer edges of the wave experience regions of greater Alfv\'en speed sooner than the central section of the wave, and so begin to refract away from them, leading to the wrapping effect first seen in \cite{mclaughlin2004mhd}. Given the radial symmetry of the Alfv\'en speed profile, it is only necessary to drive a linear fast magnetoacoustic wave toward the X-point from one orientation (say the bottom boundary, where the results for driving from the upper boundary and from the left will be identical except for a suitable axis rotation).

The introduction of gravitational stratification alters the symmetry of the Alfv\'en speed profile, which remains symmetrical in $x=0$, but is now exponentially increasing with height, as can be seen in Figures \ref{fig:2} and \ref{fig:3}. This adjustment to the symmetry of the system means that waves driven from the upper, lower, and side boundaries will all experience different regions of the Alfv\'en speed profile throughout their propagation, and so this work sets out to investigate the different behaviors, investigating waves driven from the lower (\S\ref{subsec:c1}), upper (\S\ref{sec:upper}), and left (\S\ref{sec:left}) boundaries. The right boundary is not considered, due to the symmetry in $x=0$.

Results with the linear fast magnetoacoustic wave driven from the lower boundary into the gravitationally stratified setup are fundamentally different from the stratification-free case. Figure \ref{fig:5} ($C=1$) shows that the central section of the wave propagates toward the null point more rapidly than in the $C=0$ case (Figure \ref{fig:4}), and this is a direct consequence of the exponentially increasing Alfv\'en speed with height. The key difference is the fact that the wavefront is no longer smooth, but instead piecewise smooth. This can be seen in both the numerical simulation and the WKB solution presented in Figure \ref{fig:5}. The formation of this behavior is best explained using the ray paths of the WKB solution (Figures \ref{fig:6} and \ref{fig:7}) and the ray tracing terminology discussed earlier in this section. For the blue ray paths in Figure \ref{fig:6}, which correspond to ray paths for the stratified case, multiple caustics (crossover points) are observed. These caustics, and the caustic curves connecting them, are responsible for the cusps seen in the wavefronts. Figure \ref{fig:7} demonstrates this formation. The green ray paths cross one another at caustics, and these caustics form a caustic curve. Meanwhile, the blue wavefront is clearly piecewise smooth, with two cusps, and the point of the left cusp can be seen touching the caustic curve. Moving forward in time, this behavior continues, with the location of the caustics, caustic curve, and cusp all changing, and it is clear that the caustics and caustic curves are responsible for the cusp. Figure \ref{fig:8} demonstrates the link between the value of $C$ and the location of the cusp. It is clear that increasing the level of stratification ensures that the cusps are closer to the null point for a given time. This is a consequence of the refraction away from regions of greater Alfv\'en speed, as a higher value of $C$ corresponds to a steeper exponential density profile, and hence a steeper Alfv\'en speed profile.

\cite{afanasyev2012modelling}, \cite{tarr2017magnetoacoustic}, and \cite{pennicott2021conversion} all used ray tracing methods to investigate the propagation of magnetoacoustic waves around magnetic null points, observing caustics similar to those responsible for the formation of cusps in this investigation. However, these works all consider the high-$\beta$ or nonlinear cases and as such, this paper has shown that these caustics, and the resulting cusps, can also be present in the low-$\beta$, linear case.
 
Figure \ref{fig:9} demonstrates the behavior of the linear fast magnetoacoustic wave when driven into the same setup from the upper boundary. The wave propagates toward the null point faster than in Figures \ref{fig:4} and \ref{fig:5}, and this is due to the wave being initialized into a region of higher Alfv\'en speed than was the case previously. This can be seen in Figures \ref{fig:2} and \ref{fig:3}. This leads to a tighter wrapping around the null point than in Figure \ref{fig:4}. As well as this, the outer sections of the wave exhibit different behavior to Figure \ref{fig:4}, as these sections are first experiencing the effect of refraction toward the null point, but as they escape the null point, they begin to experience the effect of refraction toward a saddle point in the Alfv\'en speed profile, located at $\left[0, -2 \right]$. This is a further demonstration that the propagation of the fast magnetoacoustic wave is dictated by the morphology of the Alfv\'en speed profile. The effect of the saddle point was not present in Figure \ref{fig:5} as in that case, the saddle point is located on the lower boundary of the simulation. Figure \ref{fig:10} highlights the effect of the saddle point, with the blue ray paths initially experiencing refraction toward the null point, before later experiencing a similar effect toward the point at $\left[0, -2 \right]$.

Finally, Figure \ref{fig:11} shows the behavior of the linear fast magnetoacoustic wave when driven into the same setup from the left boundary. The wave propagates rightward toward the null point, with the upper section of the wave propagating faster than the lower section. This is a direct consequence of the Alfv\'en speed exponentially increasing with height for a fixed value of $x$. It is clear from Figure \ref{fig:2} that the variation in Alfv\'en speed along the left boundary is around 6 orders of magnitude. This leads to the upper section of the wave reaching close to the null point and beginning to wrap around significantly quicker than the lower section. The lower section of the wave also experiences the Alfv\'en speed saddle point and this leads to the results in Figure \ref{fig:11} resembling a rotated version of the results seen in Figure \ref{fig:9}. The ray paths in Figure \ref{fig:12} also resemble those in Figure \ref{fig:10}, with a subset remaining trapped at the null point, a subset which escape the null point and experience the refraction effects toward the saddle point, and a pair of ray paths which highlight the division between the two regimes. 

This work has thus built on the understanding that the background or equilibrium Alfv\'en speed profile dictates the behavior of the linear fast magnetoacoustic wave, and adding gravitational stratification fundamentally alters the symmetry of this speed profile, thus altering the propagation of the wave. Driving different boundaries generates a wave that experiences different regions of the Alfv\'en speed profile. While the effect of wave refraction away from regions of higher Alfv\'en speed toward regions of lower Alfv\'en speed still holds true, the driving of the three different boundaries leads to three different regimes (\S \ref{sec:lower}, \S \ref{sec:upper} and \S \ref{sec:left}). In summary, there is a competition here between the effect of the Alfv\'en speed numerator - which captures the magnetic inhomogeneity - and the denominator - which captures the severity of the gravitational stratification - and which is brought together into a single background or equilibrium Alfv\'en speed profile.

Multiple recent observations of wave-null point interactions in the solar corona have been published. \cite{srivastava2019observations} presented multiwavelength SDO-AIA observations of the formation of an X-point in the solar corona, and concluded that the subsequent forced magnetic reconnection was a result of the interactions between a fast magnetoacoustic wave and the null point. \cite{kumar2024direct} presented the first EUV imaging of mode conversion near a 3D null point in the solar atmosphere. \cite{kumar2024direct} noted that their observed behavior is ``quite consistent'' with the theoretical modeling of \cite{mclaughlin2006mhd}. Theoretical understanding gained from numerical and semianalytical works has contributed to a greater understanding of observed phenomena, and the contributions of this work could help to further improve this understanding.

Forward modeling of theoretical results would significantly aid comparisons with observations. In order for this to be most effective, future work will include nonlinear effects, mode conversion, and a null point topology comparable with that of the solar atmosphere.

\begin{acknowledgments}
All authors acknowledge the UK Research and Innovation (UKRI) Science and Technology Facilities Council (STFC) for support from grant No. ST/X001008/1 and for IDL support. The research was sponsored by the DynaSun project and has thus received funding under the Horizon Europe programme of the European Union under grant agreement (no. 101131534). Views and opinions expressed are however those of the author(s) only and do not necessarily reflect those of the European Union and therefore the European Union cannot be held responsible for them. This work was also supported by the Engineering and Physical Sciences Research Council (EP/Y037464/1) under the Horizon Europe Guarantee.

R.T.S. also thanks the Royal Astronomical Society for financial assistance.
\end{acknowledgments}

%




\software{Numerical simulations were conducted with LARE2D, which is available online (https://github.com/Warwick-Plasma/Lare2d). The data that support the findings of this study are available from the corresponding author upon reasonable request. Northumbria University's Oswald high performance computing (HPC) system was used to carry out all numerical simulations in this study.}



\appendix

\section{Stratification Derivation}\label{sec:Appendix A}

The derivation of Equation (\ref{eqn:7}) begins with the MHD equations, in particular the equation of motion, the mass conservation equation, the adiabatic energy equation, and the induction equation, all four of which are given below:

\begin{equation}
   \rho \left[ \frac{\partial\textbf{v}}{\partial t} + \left(\textbf{v} \cdot \nabla \right)\textbf{v} \right] = -\nabla P + \textbf{j} \times \textbf{B} + \rho \textbf{g},
   \label{eqn:A1}
\end{equation}
\begin{equation}
   \frac{\partial \rho}{\partial t} + \nabla \cdot \left( \rho \textbf{v}\right) = 0,
   \label{eqn:A2}
\end{equation}
\begin{equation}
   \frac{\partial P}{\partial t} + \textbf{v} \cdot \nabla P = -\gamma P \left( \nabla \cdot \textbf{v} \right),
   \label{eqn:A3}
\end{equation}
\begin{equation}
   \frac{\partial \textbf{B}}{\partial t} = \nabla \times (\textbf{v} \times \textbf{B}) + \eta \nabla^2 \textbf{B}.
   \label{eqn:A4}
\end{equation}
A steady ground state ($\partial/\partial t = 0 $) and $\textbf{v} = 0$ is considered, along with the following small perturbations:

\begin{equation}
   \textbf{B} = \textbf{B}_0(\textbf{r}) + \delta \textbf{b}(\textbf{r},t),
   \label{eqn:A5}
\end{equation}
\begin{equation}
   \textbf{v} = \textbf{0} + \delta \textbf{v}_1(\textbf{r},t),
   \label{eqn:A6}
\end{equation}
\begin{equation}
  \rho = \rho_0 + \delta \rho_1(\textbf{r}, t),
  \label{eqn:A7}
\end{equation}
\begin{equation}
   P = P_0 + \delta P_1(\textbf{r}, t).
   \label{eqn:A8}
\end{equation}
As with standard perturbation theory, these perturbations are substituted into the equations, which are then expanded in powers of $\delta$, neglecting terms of order $\delta^2$ and above. Doing so yields the following linearized perturbation equations:

\begin{equation}
   \rho_0 \frac{\partial \textbf{v}_1}{\partial t} = -\nabla P_1 + \left( \frac{1}{\mu} \nabla \times \textbf{b}\right) \times \textbf{B}_0 + \rho_1 \textbf{g},
   \label{eqn:A9}
\end{equation}
\begin{equation}
   \frac{\partial \rho_1}{\partial t} = - \nabla \cdot \left( \rho_0 \textbf{v}_1\right),
   \label{eqn:A10}
\end{equation}
\begin{equation}
   \frac{\partial P_1}{\partial t} = -\gamma P_0 \left( \nabla \cdot \textbf{v}_1 \right) - \textbf{v}_1 \cdot \nabla P_0,
   \label{eqn:A11}
\end{equation}
\begin{equation}
   \frac{\partial \textbf{b}}{\partial t} =  \nabla \times \left( \textbf{v}_1 \times \textbf{B}_0 \right) + \eta \nabla^2 \textbf{b}.
   \label{eqn:A12}
\end{equation}
This paper is carried out in a 2D geometry in the $x$-$y$plane, and makes use of the following coordinate system to split the velocity field up into a parallel and a perpendicular component to the magnetic field:

\begin{equation}
   \textbf{v}_1 = \text{v}_\parallel \left(\frac{\textbf{B}_0}{\textbf{B}_0 \cdot \textbf{B}_0} \right) + \text{v}_\perp \left(\frac{ - \nabla A_0}{\textbf{B}_0 \cdot \textbf{B}_0} \right)+ v_z\hat{\textbf{z}},
   \label{eqn:A13}
\end{equation}
where $\textbf{A} = (0,0,A_0)$ is the vector potential, satisfying $\textbf{B}_0 = \nabla \times \textbf{A}$, and $\textbf{B}_0 \perp   \nabla A_0 \perp \hat{\textbf{z}}$, i.e. the system is orthogonal. Note that $\text{v}_\perp =\textbf{v} \times \textbf{B} \cdot \hat{\textbf{z}} = v_xB_y-v_yB_x$ and $\text{v}_\parallel = \textbf{v} \cdot \textbf{B} = v_xB_x+v_yB_y$. Using this coordinate system, and by  taking $\textbf{g} = \left(0, -g, 0 \right)$, the equations become

\begin{equation}
   \rho_0 \frac{\partial \text{v}_\perp}{\partial t} = -\left(\textbf{B}_0 \cdot \textbf{B}_0 \right) \left( \frac{1}{\mu} \nabla \times \textbf{b}\right) \cdot \hat{\textbf{z}} + \nabla A_0 \cdot \nabla P_1 -\rho_1 \nabla A_0 \cdot \textbf{g},
   \label{eqn:A14}
\end{equation}
\begin{equation}
   \rho_0 \frac{\partial \text{v}_\parallel}{\partial t} = -\left(\textbf{B}_0 \cdot \nabla \right) P_1 + \rho_1 \textbf{B}_0 \cdot \textbf{g},
   \label{eqn:A15}
\end{equation}
\begin{equation}
   \rho_0 \frac{\partial v_z}{\partial t} = \frac{1}{\mu}\left(\textbf{B}_0 \cdot \nabla \right) b_z,
   \label{eqn:A16}
\end{equation}
\begin{equation}
   \frac{\partial b_x}{\partial t} = \frac{\partial \text{v}_\perp}{\partial y} + \eta \nabla^2 b_x,
   \label{eqn:A17}
\end{equation}
\begin{equation}
   \frac{\partial b_y}{\partial t} = -\frac{\partial \text{v}_\perp}{\partial x} + \eta \nabla^2 b_y,
   \label{eqn:A18}
\end{equation}
\begin{equation}
   \frac{\partial b_z}{\partial t} = \left( \textbf{B}_0 \cdot \nabla \right) v_z + \eta \nabla^2 b_z,
   \label{eqn:A19}
\end{equation}
\begin{equation}
   \frac{\partial P_1}{\partial t} = v_y\rho_0g-\gamma P_0 \left[ \nabla \cdot \left( \frac{\textbf{B}_0 \text{v}_\parallel}{\textbf{B}_0 \cdot \textbf{B}_0} \right) - \nabla \cdot \left( \frac{\text{v}_\perp \nabla A_0}{\textbf{B}_0 \cdot \textbf{B}_0} \right) \right],
   \label{eqn:A20}
\end{equation}
\begin{equation}
   \frac{\partial \rho_1}{\partial t} = - \left[ \nabla \cdot \left( \frac{\rho_0\textbf{B}_0 \text{v}_\parallel}{\textbf{B}_0 \cdot \textbf{B}_0} \right) - \nabla \cdot \left( \frac{\rho_0\text{v}_\perp \nabla A_0}{\textbf{B}_0 \cdot \textbf{B}_0} \right) \right].
   \label{eqn:A21}
\end{equation}
This system of equations is nondimensionalized using the following: $\text{v}_\perp = \bar{B} \bar{v}\text{v}_\perp^*$, $\text{v}_\parallel = \bar{B} \bar{v}\text{v}_\parallel^*$, $\textbf{v} = \bar{v}\textbf{v}^*$, $\textbf{B}_0 = \bar{B}\textbf{B}_0^*$, $\textbf{b} = \bar{B} \textbf{b}^*$, $x = \bar{L}x^*$, $y = \bar{L}y^*$, $z = \bar{L}z^*$, $P_0 = \bar{P} P_0^*$, $P_1 = \bar{P} P_1^*$, $\nabla = \left(1/\bar{L}\right)\nabla^*$, $t=\bar{t} t^*$, $A_0 = \bar{B} \bar{L} A_0^*$, $\rho_0 = \bar{\rho}\rho_0^*$, $\rho_1 = \bar{\rho}\rho_1^*$, $\textbf{g} = \bar{g}\textbf{g}^*$ and $\eta = \bar{\eta}$ where $^*$ denotes the dimensionless quantities and $\bar{}$ denotes constants with the dimensions of the variable they are scaling. The following relations are also set: $\bar{v} = \bar{B} / \sqrt{\mu \bar{\rho}}$, $\bar{t} = \bar{L}/\bar{v}$, $\bar{g} = \bar{L}/\bar{t}^2$, $R_m^{-1} = \bar{\eta} \bar{t}/\bar{L}^2$ and $\bar{\beta} = 2\mu \bar{P}/\bar{B}^2$.

After applying this nondimensionalization, the system of equations becomes

\begin{equation}
    \rho_0^* \frac{\partial \text{v}_\perp^*}{\partial t^*} = - \left(\textbf{B}_0^*\cdot \textbf{B}_0^*\right)\left(\nabla^*\times\textbf{b}^* \right) \cdot \hat{\textbf{z}} + \frac{\bar{\beta}}{2} \nabla^*A_0^*\cdot\nabla^*P_1^* - \rho_1^*\nabla^*A_0^*\cdot\textbf{g}^*,
    \label{eqn:A22}
\end{equation}
\begin{equation}
    \rho_0^* \frac{\partial \text{v}_\parallel^*}{\partial t^*} = - \frac{\bar{\beta}}{2}\left( \textbf{B}_0^* \cdot \nabla^* \right)P_1^* + \rho_1^*\textbf{B}_0^*\cdot \textbf{g}^*,
    \label{eqn:A23}
\end{equation}
\begin{equation}
    \rho_0^* \frac{\partial v_z^*}{\partial t^*} = \left(\textbf{B}_0^*\cdot\nabla^* \right)b_z^*,
    \label{eqn:A24}
\end{equation}
\begin{equation}
    \frac{\partial b_x^*}{\partial t^*} = \frac{\partial\text{v}_\perp^*}{\partial y^*} + \frac{1}{R_m}\nabla^{*^2}b_x^*,
    \label{eqn:A25}
\end{equation}
\begin{equation}
    \frac{\partial b_y^*}{\partial t^*} = -\frac{\partial\text{v}_\perp^*}{\partial x^*} + \frac{1}{R_m}\nabla^{*^2}b_y^*,
    \label{eqn:A26}
\end{equation}
\begin{equation}
    \frac{\partial b_z^*}{\partial t^*} = \left(\textbf{B}_0^* \cdot \nabla^*\right)v_z^* + \frac{1}{R_m}\nabla^{*^2}b_z^*,
    \label{eqn:A27}
\end{equation}
\begin{equation}
    \frac{\bar{\beta}}{2}\frac{\partial P_1^*}{\partial t^*} = v_y^*\rho_0^*g^*- \frac{\bar{\beta}}{2}\gamma P_0^* \left[\nabla^* \cdot \left( \frac{\textbf{B}_0^*\text{v}_\parallel^*}{\textbf{B}_0^*\cdot \textbf{B}_0^*}\right) - \nabla^* \cdot \left( \frac{\text{v}_\perp^*\nabla^* A_0^*}{\textbf{B}_0^*\cdot \textbf{B}_0^*}\right) \right],
    \label{eqn:A28}
\end{equation}
\begin{equation}
    \frac{\partial \rho_1^*}{\partial t^*} = - \left[\nabla^* \cdot \left( \frac{\rho_0^*\textbf{B}_0^*\text{v}_\parallel^*}{\textbf{B}_0^*\cdot \textbf{B}_0^*}\right) - \nabla^* \cdot \left( \frac{\rho_0^*\text{v}_\perp^*\nabla^* A_0^*}{\textbf{B}_0^*\cdot \textbf{B}_0^*}\right) \right].
    \label{eqn:A29}
\end{equation}
Considering an ideal plasma ($\eta = 0$ or $R_m \rightarrow \infty$), and omitting the $^*$ notation yields the following system of equations:

\begin{equation}
    \rho_0 \frac{\partial \text{v}_\perp}{\partial t} = - \left(\textbf{B}_0\cdot \textbf{B}_0\right)\left(\nabla\times\textbf{b} \right) \cdot \hat{\textbf{z}} + \frac{\bar{\beta}}{2} \nabla A_0\cdot\nabla P_1  - \rho_1\nabla A_0\cdot\textbf{g},
    \label{eqn:A30}
\end{equation}
\begin{equation}
    \rho_0 \frac{\partial \text{v}_\parallel}{\partial t} = - \frac{\bar{\beta}}{2}\left( \textbf{B}_0 \cdot \nabla \right)P_1 + \rho_1\textbf{B}_0\cdot \textbf{g},
    \label{eqn:A31}
\end{equation}
\begin{equation}
    \rho_0 \frac{\partial v_z}{\partial t} = \left(\textbf{B}_0\cdot\nabla \right)b_z,
    \label{eqn:A32}
\end{equation}
\begin{equation}
    \frac{\partial b_x}{\partial t} = \frac{\partial\text{v}_\perp}{\partial y},
    \label{eqn:A33}
\end{equation}
\begin{equation}
    \frac{\partial b_y}{\partial t} = -\frac{\partial\text{v}_\perp}{\partial x},
     \label{eqn:A34}
\end{equation}
\begin{equation}
    \frac{\partial b_z}{\partial t} = \left(\textbf{B}_0 \cdot \nabla\right)v_z,
    \label{eqn:A35}
\end{equation}
\begin{equation}
    \frac{\bar{\beta}}{2}\frac{\partial P_1}{\partial t} = v_y\rho_0g- \frac{\bar{\beta}}{2}\gamma P_0 \left[\nabla \cdot \left( \frac{\textbf{B}_0\text{v}_\parallel}{\textbf{B}_0\cdot \textbf{B}_0}\right) - \nabla \cdot \left( \frac{\text{v}_\perp\nabla A_0}{\textbf{B}_0\cdot \textbf{B}_0}\right) \right],
    \label{eqn:A36}
\end{equation}
\begin{equation}
    \frac{\partial \rho_1}{\partial t} = - \left[\nabla \cdot \left( \frac{\rho_0\textbf{B}_0\text{v}_\parallel}{\textbf{B}_0\cdot \textbf{B}_0}\right) - \nabla \cdot \left( \frac{\rho_0\text{v}_\perp\nabla A_0}{\textbf{B}_0\cdot \textbf{B}_0}\right) \right].
    \label{eqn:A37}
\end{equation}
Considering $\textbf{B}_0 = \left(B_x, B_y, 0 \right)$ and $\textbf{b} = \left(b_x, b_y, b_z \right)$ allows the equations governing the $z$-direction to be isolated. Equations (\ref{eqn:A32}) and (\ref{eqn:A35}) can be combined to give a single wave equation for the linear Alfv\'en wave:

\begin{equation}
    \rho_0 \frac{\partial^2 v_z}{\partial t^2} = \left(B_x \frac{\partial}{\partial x} + B_y \frac{\partial}{\partial y}\right)^2v_z.
    \label{eqn:A38}
\end{equation}
Similarly, Equations (\ref{eqn:A30}), (\ref{eqn:A31}), (\ref{eqn:A36}) and (\ref{eqn:A37}) become

\begin{equation}
    \rho_0 \frac{\partial \text{v}_\perp}{\partial t} = - \left(B_x^2+B_y^2\right)\left( \frac{\partial b_y}{\partial x} - \frac{\partial b_x}{\partial y}\right)  + \frac{\bar{\beta}}{2} \nabla A_0 \cdot \nabla P_1  + \rho_1B_xg,
    \label{eqn:A39}
\end{equation}
\begin{equation}
    \rho_0 \frac{\partial \text{v}_\parallel}{\partial t} = - \frac{\bar{\beta}}{2}\left( \textbf{B}_0 \cdot \nabla\right)P_1 - \rho_1B_yg,
    \label{eqn:A40}
\end{equation}
\begin{equation}
    \frac{\bar{\beta}}{2}\frac{\partial P_1}{\partial t} = v_y\rho_0g- \frac{\bar{\beta}}{2}\gamma P_0 \left[\nabla \cdot \left( \frac{\textbf{B}_0\text{v}_\parallel}{B_x^2+B_y^2}\right) - \nabla \cdot \left( \frac{\text{v}_\perp\nabla A_0}{B_x^2+B_y^2}\right) \right],
    \label{eqn:A41}
\end{equation}
\begin{equation}
    \frac{\partial \rho_1}{\partial t} = - \left[\nabla \cdot \left( \frac{\rho_0\textbf{B}_0\text{v}_\parallel}{B_x^2+B_y^2}\right) - \nabla \cdot \left( \frac{\rho_0\text{v}_\perp\nabla A_0}{B_x^2+B_y^2}\right) \right],
    \label{eqn:A42}
\end{equation}
while Equations (\ref{eqn:A33}) and (\ref{eqn:A34}) remain the same. Taking the time derivatives of Equations (\ref{eqn:A39}) and (\ref{eqn:A40}), and combining with Equations (\ref{eqn:A33}) and (\ref{eqn:A34}) yields the following two equations:

\begin{equation}
    \rho_0 \frac{\partial^2 \text{v}_\perp}{\partial t^2} = \left(B_x^2+B_y^2\right)\left( \frac{\partial^2}{\partial x^2} + \frac{\partial^2}{\partial y^2}\right)\text{v}_\perp  + \frac{\bar{\beta}}{2} \nabla A_0 \cdot \nabla \frac{\partial P_1}{\partial t}  + B_xg\frac{\partial \rho_1}{\partial t},
    \label{eqn:A43}
\end{equation}
\begin{equation}
    \rho_0 \frac{\partial^2 \text{v}_\parallel}{\partial t^2} = - \frac{\bar{\beta}}{2}\left( \textbf{B}_0 \cdot \nabla\right)\frac{\partial P_1}{\partial t} - B_yg \frac{\partial \rho_1}{\partial t},
    \label{eqn:A44}
\end{equation}
where the time derivatives of $P_1$ and $\rho_1$ are defined in Equations (\ref{eqn:A41}) and (\ref{eqn:A42}), respectively. Substituting these derivatives into Equations (\ref{eqn:A43}) and (\ref{eqn:A44}) yields the following two equations:

\begin{multline}
    \rho_0 \frac{\partial^2 \text{v}_\perp}{\partial t^2} = \left(B_x^2+B_y^2\right)\left( \frac{\partial^2}{\partial x^2} + \frac{\partial^2}{\partial y^2}\right)\text{v}_\perp 
    \\ + \frac{\bar{\beta}}{2} \nabla A_0 \cdot \nabla \left[ \frac{2}{\bar{\beta}}\rho_0g \left( \frac{\text{v}_\parallel B_y - \text{v}_\perp B_x}{B_x^2+B_y^2} \right) - \gamma P_0 \left[\nabla \cdot \left( \frac{\textbf{B}_0\text{v}_\parallel}{B_x^2+B_y^2}\right) - \nabla \cdot \left( \frac{\text{v}_\perp\nabla A_0}{B_x^2+B_y^2}\right) \right] \right] \\ - B_x g \left[\nabla \cdot \left( \frac{\rho_0\textbf{B}_0\text{v}_\parallel}{B_x^2+B_y^2}\right) - \nabla \cdot \left( \frac{\rho_0\text{v}_\perp\nabla A_0}{B_x^2+B_y^2}\right) \right],
        \label{eqn:A45}
\end{multline}
\begin{multline}
    \rho_0 \frac{\partial^2 \text{v}_\parallel}{\partial t^2} = - \frac{\bar{\beta}}{2}\left( \textbf{B}_0 \cdot \nabla\right) \left[ \frac{2}{\bar{\beta}}\rho_0g \left( \frac{\text{v}_\parallel B_y - \text{v}_\perp B_x}{B_x^2+B_y^2} \right) - \gamma P_0 \left[\nabla \cdot \left( \frac{\textbf{B}_0\text{v}_\parallel}{B_x^2+B_y^2}\right) - \nabla \cdot \left( \frac{\text{v}_\perp\nabla A_0}{B_x^2+B_y^2}\right) \right] \right] \\
    + B_y g \left[\nabla \cdot \left( \frac{\rho_0\textbf{B}_0\text{v}_\parallel}{B_x^2+B_y^2}\right) - \nabla \cdot \left( \frac{\rho_0\text{v}_\perp\nabla A_0}{B_x^2+B_y^2}\right) \right].
        \label{eqn:A46}
\end{multline}

\section{Vertical and Horizontal Field Examples: Agreement with the Literature}\label{sec:Appendix B}

Equations (\ref{eqn:A45}) and (\ref{eqn:A46}) are valid for any choice of magnetic field, and two logical choices of magnetic field to investigate are first that of a purely vertical magnetic field, and second a purely horizontal magnetic field. This appendix covers the introduction of those two magnetic field choices, and how Equations (\ref{eqn:A45}) and (\ref{eqn:A46}) reduce to Equations (9.111), (9.126), and (9.128) of \cite{roberts2019mhd}.

Under the case of a purely vertical magnetic field, with $B_x = 0$, $B_y=B_0$, $\partial/\partial x=0$ and subsequently $\text{v}_\bot = B_0v_x$, $\text{v}_\parallel = B_0v_y$ and $\nabla A_0 = -B_0 \hat{\textbf{x}}$, Equation (\ref{eqn:A45}) reduces to

\begin{equation}
    \frac{\partial^2v_x}{\partial t^2} = v_A^2(y)\frac{\partial^2v_x}{\partial y^2},
    \label{eqn:B47}
\end{equation}
and similarly Equation (\ref{eqn:A46}) reduces to

\begin{equation}
    \frac{\partial^2v_y}{\partial t^2} = c_s^2(y) \frac{\partial^2 v_y}{\partial t^2} + \frac{\gamma}{\rho_0}\frac{\partial P_0}{\partial y} \frac{\partial v_y}{\partial y}.
    \label{eqn:B48}
\end{equation}
Transforming to the coordinate system used in \cite{roberts2019mhd}, i.e. $y\rightarrow -z$ yields both halves of their Equation (9.111). Note that they choose their $z$-axis to be pointing downwards, hence this paper has $\partial P_0/\partial y = - \rho_0g$ but \cite{roberts2019mhd} has $\partial P_0/\partial z = \rho_0g$. 

Under the case of a purely horizontal magnetic field, with $B_x = B_0(y)$, $B_y = 0$, $\Delta = \partial v_x/\partial x + \partial v_y/\partial y$ and subsequently $\text{v}_\bot = -B_0v_y$, $\text{v}_\parallel = B_0v_x$ and $\nabla A_0 = B_0(y) \hat{\textbf{y}}$, Equation (\ref{eqn:A45}) reduces to:

\begin{equation}
    \frac{\partial^2v_y}{\partial t^2} = v_A^2(y) \left(\frac{\partial^2}{\partial x^2} + \frac{\partial^2}{\partial y^2}\right)v_y + \left(g + \frac{2B_0B_0'}{\rho_0} \right)\frac{\partial v_y}{\partial y} + c_s^2(y) \frac{\partial \Delta}{\partial y} + \frac{\gamma\Delta}{\rho_0}\frac{\partial P_0}{\partial y} + g \left( \frac{\partial v_x}{\partial x} - \frac{\partial v_y}{\partial y} \right),
    \label{eqn:B49}
\end{equation}
and similarly Equation (\ref{eqn:A46}) reduces to:

\begin{equation}
    \frac{\partial^2 v_x}{\partial t^2} = c_s^2(y) \frac{\partial^2 v_x}{\partial x^2} - \left[g - c_s^2(y) \frac{\partial}{\partial y} \right] \frac{\partial v_y}{\partial x}.
    \label{eqn:B50}
\end{equation}
Transforming to the coordinate system of \cite{roberts2019mhd} in the same way as before yields both their Equations (9.126) and (9.128).

\section{Simplifications and WKB Formulation}\label{sec:Appendix C}

When considering the nondimensionalizing setup in the context of the solar corona, a small $\bar{P}$ is chosen. Given that $\bar{\beta} = 2\mu\bar{P}/\bar{B}^2$, it is clear that $\bar{\beta} \ll 1$, except at the null point. Note that the null point is a consequence of the choice of equilibrium magnetic field $\textbf{B}_0=\left(x, -y, 0\right)$ given in Equation (\ref{eqn:5}).

Similarly, given that $\bar{P} \ll 1$ and $\bar{P} = k_B \bar{\rho} \bar{T}$, it follows that $\bar{T} \ll 1$ (i.e. the cold plasma approximation). Given this fact, and that $\bar{\beta} = 1$ occurs very close to the null point, it is reasonable to neglect the $\bar{\beta}$ terms in the derivation of the WKB solution.

However, care must be taken when making this consideration. The density profile given in Equation (\ref{eqn:6}) and used throughout the paper contains $C$, where $C= g / 2T$. It is clear that for a finite $C$ (in the case of this paper $C=1$), for $T \ll 1$, it follows that $g \ll 1$, and as such any term with a $g$ coefficient can be omitted.

The choice of equilibrium magnetic field, given in Equation (\ref{eqn:5}) by $\textbf{B}_0 = \left( x, -y, 0\right)$, can be substituted into Equation (\ref{eqn:A45}) and doing so, along with omitting terms with $\bar{\beta}$ and $g$ coefficients, yields the following equation:

\begin{equation}
    \frac{\partial^2 \text{v}_\perp}{\partial t^2} = \frac{1}{\rho_0}\left(B_x^2+B_y^2\right)\left( \frac{\partial^2}{\partial x^2} + \frac{\partial^2}{\partial y^2}\right)\text{v}_\perp = \frac{1}{\rho_0}\left(x^2+y^2\right)\left( \frac{\partial^2}{\partial x^2} + \frac{\partial^2}{\partial y^2}\right)\text{v}_\perp.
    \label{eqn:C51}
\end{equation}
These simplifications can be confirmed as suitable by redimensionalizing Equation (\ref{eqn:A45}) and introducing typical coronal values to the equation. Doing so allows an order of magnitude argument to be made - the first term on the right-hand side of Equation (\ref{eqn:A45}) is 8 orders of magnitude larger then the other two terms, and as such, Equation (\ref{eqn:C51}) is a suitable governing equation for the system.

From Equation (\ref{eqn:C51}), the WKB solution can be derived. The following form of $\text{v}_\perp$ is substituted into Equation (\ref{eqn:C51}): $\text{v}_\perp = \text{exp}\left(i\left[\phi\left(x,y\right)-\omega t\right]\right)$, as well as the equilibrium magnetic field and density profile used throughout this paper. Doing so, defining $p = \partial\phi/\partial x$ and $q=\partial \phi /\partial y$ and making the WKB approximation $\left( \phi \sim \omega \gg 1 \right)$, followed by using Charpit’s method and working through the necessary algebra yields the final system of five equations, which are solved using a fourth-order Runge-Kutta method to give WKB solutions for the propagation of the fast magnetoacoustic wave:

\begin{equation}
    \frac{d\phi}{ds}= \omega^2,
    \label{eqn:C52}
\end{equation}
\begin{equation}
    \frac{dx}{ds} = e^{C\left(y+2\right)}\left(x^2 + y^2\right)p,
    \label{eqn:C53}
\end{equation}
\begin{equation}
    \frac{dy}{ds} = e^{C\left(y+2\right)}\left(x^2 + y^2\right)q,
    \label{eqn:C54}
\end{equation}
\begin{equation}
    \frac{dp}{ds} =- e^{C\left(y+2\right)}\left(p^2 + q^2\right)x,
    \label{eqn:C55}
\end{equation}
\begin{equation}
    \frac{dq}{ds} =- e^{C\left(y+2\right)}\left(p^2 + q^2\right)\left[y + \frac{C}{2}\left(x^2 + y^2\right)\right].
    \label{eqn:C56}
\end{equation}


\bibliography{sample631}{}
\bibliographystyle{aasjournal}



\end{document}